\documentclass[
reprint,
superscriptaddress,
showpacs,
preprintnumbers,
nofootinbib,
nobibnotes,
amsmath,
amssymb, 
aps,
prd,
floatfix
]{revtex4-1}

\usepackage{xcolor}
\usepackage[normalem]{ulem}

\usepackage{fontawesome5}

\usepackage[export]{adjustbox}
\usepackage[colorlinks=true,breaklinks=true]{hyperref}
\hypersetup{urlcolor=blue,
	    citecolor=black}

\usepackage{graphicx}% Include figure files
\usepackage{dcolumn}% Align table columns on decimal point
\usepackage{bm}% bold math
\usepackage{amsmath, amssymb}
\usepackage{graphicx}
\usepackage{multirow}
\usepackage{float}
\usepackage{physics}

\def\be{\begin{equation}}
\def\ee{\end{equation}}

\newcommand{\dPBH}{{d}_{\rm PBH}}

\newcommand{\GeV}{\,\text{GeV}}

\newcommand{\al}[1]{\begin{align}\begin{aligned} #1 \end{aligned}\end{align}}

\begin{document}

\preprint{IFT-UAM/CSIC-25-58}

\title{Chasing  Serendipity: Tackling Transient Sources with  Neutrino Telescopes}

\author{Lua F. T. Airoldi}
\email{lua.airoldi@usp.br}
\affiliation{Instituto de F\'isica, Universidade de S\~ao Paulo, C.P. 66.318, 05315-970 S\~ao Paulo, Brazil}

\author{Gustavo F. S. Alves}%
\email{gustavo.figueiredo.alves@usp.br}
\affiliation{Instituto de F\'isica, Universidade de S\~ao Paulo, C.P. 66.318, 05315-970 S\~ao Paulo, Brazil}

\author{Yuber F. Perez-Gonzalez}%
\email{yuber.perez@uam.es}
\affiliation{Departamento de Física Teórica and Instituto de Física Teórica UAM/CSIC,Universidad Autónoma de Madrid, Cantoblanco, 28049 Madrid, Spain}

\author{Gabriel M. Salla}%
\email{gabriel.massoni.salla@usp.br}
\affiliation{Instituto de F\'isica, Universidade de S\~ao Paulo, C.P. 66.318, 05315-970 S\~ao Paulo, Brazil}

\author{Renata Zukanovich Funchal}%
\email{zukanov@if.usp.br}
\affiliation{Instituto de F\'isica, Universidade de S\~ao Paulo, C.P. 66.318, 05315-970 S\~ao Paulo, Brazil}
%  \altaffiliation[Also at ]{Physics Department, XYZ University.}%Lines break automatically or can be forced with \\

\date{\today}

\begin{abstract}

The discovery of ultra-high-energy neutrinos by IceCube marked the beginning of neutrino astronomy. Yet, the origin and production mechanisms of these neutrinos remain an open question. 
With the observation of several neutrino events with energies about the PeV, transient sources---astrophysical objects that emit particles in brief, localized bursts---have emerged as promising candidates.
In this work, we revisit the identification of such sources in IceCube and future neutrino telescopes, focusing on how both the timing and sky localization of the source affect the detection sensitivity. 
We present a framework to account for the source's right ascension in determining the effective area of detectors not located at the poles, such as KM3NeT.
As a case study, we investigate evaporating primordial black holes (PBHs) as transient neutrino sources, showing that the detection prospects and localization accuracy are strongly influenced by the PBH’s position in the sky. Our results emphasize the complementarity between neutrino and gamma-ray observatories and showcase the potential of a global network of neutrino detectors to identify and localize transient events that might be missed by traditional photon-based instruments.

\end{abstract}

%\keywords{Suggested keywords}%Use showkeys class option if keyword
                              %display desired
\maketitle

%\tableofcontents

\section{\label{sec:intro}Introduction}

The landmark detection of the first ultra-high-energy neutrino events by IceCube---boasting energies around 1 PeV---marked a monumental breakthrough, unveiling particles that had journeyed from beyond our galaxy~\cite{IceCube:2013cdw,IceCube:2013low}, heralds the dawn of a new era of neutrino astronomy, started by the first observation of solar neutrinos~\cite{Davis:1968cp}. Since then, IceCube was able to measure hundreds of similar events during the last ten years, thus characterizing a diffuse flux of neutrinos from astrophysical origin~\cite{IceCube:2014stg,Silva:2023wol,IceCube:2023hou}. While the extragalactic astrophysical neutrino flux has been measured by IceCube with high significance, the actual production mechanism for super TeV neutrinos is still an open field of research~\cite{Vitagliano:2019yzm,Ahlers:2018fkn,Kurahashi:2022utm,Groth:2025aan}. Among possible candidates, there are transient sources that, contrary to steady ones, emit neutrinos only during a very narrow time window. Examples of known transient sources are gamma-ray bursts~\cite{IceCube:2017amx,Abbasi:2022whi,IceCube:2023woj}, binary mergers~\cite{Kurahashi:2022utm,Zhou:2023rtr} and supernovae~\cite{IceCube:2023esf}, but other novel sources can also be considered, e.g. evaporating primordial black holes~\cite{Carr:2009jm,Carr:2020gox,Carr:2020xqk,Green:2020jor,Khlopov:2008qy,DeRomeri:2024zqs,Capanema:2021hnm}. The search for such short-duration signals is challenging. It demands the ability to identify localized bursts of neutrino events against the ever-present diffuse background flux~\cite{Lucarelli:2022moc}. Equally vital is the precise localization of the source in the sky, as it determines the neutrino's trajectory through the Earth---affecting both the detection probability and the interpretation of the event.

Recently, the KM3NeT collaboration made a groundbreaking observation of a neutrino-like event at the astonishing energy  of $\sim$1 EeV, a detection that eluded IceCube. This remarkable finding has ignited a wave of excitement within the scientific community, underscoring the crucial importance of deploying multiple neutrino telescopes across diverse global locations.
While the origin of the KM3-230213A event remains uncertain, a point-like transient source is a plausible explanation, one that could even help reconcile this observation with the absence of a corresponding signal in IceCube~\cite{Li:2025tqf,Neronov:2025jfj}. For more discussions on the possible origin of the KM3-230213A event, see Refs.~\cite{Anchordoqui:2025xug, Dev:2025czz, Farzan:2025ydi, Baker:2025cff, Zhang:2025rqh, He:2025bex, Khan:2025gxs, Murase:2025uwv, Das:2025vqd, Zhang:2025abk, Barman:2025hoz, Choi:2025hqt, Dvali:2025ktz, Jho:2025gaf, Klipfel:2025jql, Jiang:2025blz, Crnogorcevic:2025vou, Alves:2025xul, Wang:2025lgn, Narita:2025udw, Kohri:2025bsn, Brdar:2025azm, Borah:2025igh, Boccia:2025hpm, Yang:2025kfr, Neronov:2025jfj, Dzhatdoev:2025sdi, Satunin:2025uui, Muzio:2025gbr}. This example serves to highlight the indispensable role of neutrino telescopes, as so far no accompanying electromagnetic signals have been detected~\cite{Das:2025vqd, Crnogorcevic:2025vou, Wang:2025lgn, Filipovic:2025ulm, Narita:2025udw, Neronov:2025jfj, Dzhatdoev:2025sdi, Fang:2025nzg, Muzio:2025gbr}. This absence may be attributed to several factors: the burst may have occurred outside the field of view of the detector, the detector was not operating at that moment, the photon energy may have fallen below the threshold of detection, or it is entirely possible that no such emissions were generated at all. In what follows, we show how a coordinated effort among multiple neutrino telescopes is essential in such cases.

Our main goal in this paper is to revisit the identification of transient sources in IceCube and in future neutrino telescopes. In particular, we are interested in studying the impact of the time and localization dependence of the source on the detection sensitivity. For IceCube, due to its privileged position in the South Pole, the number of events is parametrized by an effective area~\cite{IceCube:2021xar} $A_\text{eff}^\text{IC}(E,\delta)$ that is a function solely of the neutrino energy $E$ and its declination $\delta$ (corresponding to a certain latitude). For current and future neutrino telescopes, for instance KM3NeT~\cite{KM3Net:2016zxf}, that are not located at either the South or the North Poles, both the source declination and right ascension RA (related to a longitude) become relevant. As a result, the effective area for these experiments will be functions of both variables, $A_\text{eff}(E,\delta,\text{RA})$. Intuitively, for a given time and declination, certain values of RA correspond to neutrinos arriving from above the detector. In these cases, the neutrinos are not attenuated by passage through the Earth, and the effective area is significantly 
enhanced~\footnote{This is exactly the case for events coming in shallow down-going angles. In general, 
 events  arriving from above in both experiments will experience some attenuation due to the limited overburden: the 1.5 km of ice for IceCube or the 2.3 km of sea water for KM3NeT.}. For other RA values, the neutrinos must traverse the Earth, leading to absorption or attenuation and consequently a reduced $A_\text{eff}$. 
In what follows, we present a systematic framework to incorporate the RA dependence into the effective area. This approach enables consistent and realistic searches for transient sources across the future global network of neutrino telescopes, particularly for short-duration bursts.

We will also emphasize the unique complementarity between neutrino telescopes and gamma-ray observatories in the pursuit of transient sources. While gamma-ray telescopes are powerful tools, they are constrained by a limited field of view and restricted observational time per year. Neutrino detectors, by contrast, continuously monitor the entire sky, making them invaluable partners in the search for transient cosmic phenomena.

As an application of our results, we consider the case of evaporating primordial black holes (PBH). Through Hawking radiation~\cite{Hawking:1975vcx}, a PBH can emit all sorts of particles while losing mass and, as it draws near to the complete depletion of its mass as it increases its temperature, the production of high-energy particles quickly spikes, leading to a burst of particles in a short period of time~\cite{Hawking:1974rv,Ukwatta:2015iba,Auffinger:2022khh,Baker:2021btk}.
The scenario we will analyze is that of a single PBH inside the solar system at a distance $\dPBH$ from Earth and located in the sky at $(\delta_\text{PBH},\text{RA}_\text{PBH})$. 
We will revisit how to compute the number of events coming from a PBH in IceCube and other neutrino telescopes (for previous analyses see Refs.~\cite{Halzen:1995hu,DeRomeri:2024zqs,Capanema:2021hnm,Perez-Gonzalez:2023uoi,Dave:2019epr,Dasgupta:2019cae}) while taking into account the effects of $\text{RA}_\text{PBH}$. 
Our results show that the number of events recorded by each detector is extremely dependent on the localization of both the experiment on Earth and also of the PBH, hence each neutrino telescope can provide a different amount of information for distinct values of $(\delta_\text{PBH},\text{RA}_\text{PBH})$. 
The application of the methodology developed in this work to test the hypothesis that the KM3-230213A event originated from a PBH burst near Earth is presented in the companion paper, Ref.~\cite{Airoldi:2025opo}.

The paper is organized as follows. In Section~\ref{sec:point-like} we introduce point-like transient sources and discuss the complementarity between photon and neutrino detectors. In particular, we highlight that gamma-ray experiments might miss transient sources. Then, in Section~\ref{sec:Aeff} we introduce our methodology to compute the effective area of each neutrino telescope while keeping the dependence on the right ascension of the source. With these results we show in Section~\ref{sec:PBHs} how to implement our approach to the search for exploding PBHs and we conclude in Section~\ref{sec:conclusions}.
We include Appendix~\ref{app:burst} where we describe in more detail the details of the PBH's time evolution and the event rate associated to it, and Appendix~\ref{app:triang}, where we briefly describe an additional, although significantly less powerful, tool to determine the burst's location in the sky.

%%%%%%%%%%%%%%%%%%%%%%%%%%%%%%%%%%%%%%%%%%%%%%%%%%%%%%%%%%%%%%%%%%%%%%%%%%%%%%%%%%%%%%%%%%%%%%%%%%%%%%%%%%%%%%%%%%%%%%%%%%%%%%%%%%%%%%%%%%%%%%%%%%%%%%%%%

\section{\label{sec:point-like}Point-like transient sources}

Point-like transient sources are astrophysical objects that increase their brightness only within a short time window. Due to their typically large distances from Earth, they appear as points in the sky. Among the particles they emit, high-energy neutrinos and photons are the most likely to reach detectors on Earth. Well-established examples of transient sources include gamma-ray bursts~\cite{IceCube:2017amx,Abbasi:2022whi,IceCube:2023woj}, binary mergers~\cite{Kurahashi:2022utm} and supernov\ae\ explosions~\cite{IceCube:2023esf}. In addition to these, more speculative scenarios can also be considered, e.g. the final burst of a PBH~\cite{Hawking:1974rv,Ukwatta:2015iba,Auffinger:2022khh,Baker:2021btk}. The following discussion applies to any transient source, speculative or not, that emits particles in bursts lasting for hundreds of seconds. 

Neutrino telescopes are well-suited for searching for transient signals, as neutrinos travel undisturbed through space, preserving directional information that can be reconstructed from the tracks muons leave when muon neutrinos interact with the detector material. These instruments also benefit from a large field of view, covering at least $2\pi$ of the sky, depending on the neutrino energy, and their large detection volumes compensate for the small neutrino interaction cross-sections. These advantages can be further enhanced by combining data from multiple neutrino observatories, effectively increasing the detection probability and the portion of the sky that can be monitored simultaneously, enabling nearly a full-sky coverage~\cite{Schumacher:2025qca}. As we will discuss in more detail later, 
the coordinated operation of neutrino telescopes significantly amplifies our capacity to detect and precisely localize transient astrophysical sources, offering enhanced sensitivity and greater confidence in identifying rare, high-energy cosmic events. To illustrate the benefits of this synergy, we will consider the currently operating IceCube detector~\cite{IceCube:2016zyt} and its future upgrade IceCube-Gen2~\cite{IceCube-Gen2:2020qha}, the under-construction experiments KM3NeT~\cite{KM3Net:2016zxf} and Baikal-GVD~\cite{Baikal-GVD:2018isr}, as well as the planned detectors P-ONE~\cite{P-ONE:2020ljt} and TRIDENT~\cite{TRIDENT:2022hql} (see Table~\ref{tab:Experiments_Location}).
\begin{table}[t!]
    \def\arraystretch{1.5}
        \begin{tabular}{|c|c|c|c|c|c|}
        \hline
            \textcolor{white}{aaa} & \textbf{Experiment} & \textbf{Latitude} & \textbf{Longitude} & \textbf{\begin{tabular}[c]{@{}c@{}}Volume\\ scaling\end{tabular}} & \textbf{\begin{tabular}[c]{@{}c@{}}FoV\\ angle\end{tabular}} \\ \hline
            \multirow{6}{*}{$\nu$} & IceCube & $-90^\circ$ & $0^\circ$ & 1.0 & \multirow{6}{*}{-} \\ \cline{2-5}
             & IceCube-Gen2 & $-90^\circ$ & $0^\circ$ & 7.5 &  \\ \cline{2-5}
             & KM3NeT & $36.26^\circ$ & $16.1^\circ$ & 1.0 &  \\ \cline{2-5}
             & Baikal-GVD & $53.55^\circ$ & $108.16^\circ$ & 1.0 &  \\ \cline{2-5}
             & P-ONE & $47.74^\circ$ & $-127.72^\circ$ & 1.0 &  \\ \cline{2-5}
             & TRIDENT & $17.4^\circ$ & $114^\circ$ & 7.5 &  \\ \hline\hline
            \multirow{12}{*}{$\gamma$} & LHAASO & $29.36^\circ$ & $100.14^\circ$ & \multirow{12}{*}{-} & $60^\circ$ \\ \cline{2-4} \cline{6-6} 
             & HAWC & $19^\circ$ & $-97.3^\circ$ &  & $50^\circ$ \\ \cline{2-4} \cline{6-6} 
             & VERITAS & $31.67^\circ$ & $-110.95^\circ$ &  & $3.5^\circ$ \\ \cline{2-4} \cline{6-6} 
             & HESS & $-23.27^\circ$ & $16.5^\circ$ &  & $5^\circ$ \\ \cline{2-4} \cline{6-6} 
             & \begin{tabular}[c]{@{}c@{}}Telescope\\ Array\end{tabular} & $39.3^\circ$ & $-112.9^\circ$ &  & $45^\circ$ \\ \cline{2-4} \cline{6-6} 
             & \begin{tabular}[c]{@{}c@{}}Pierre\\ Auger\end{tabular} & $-35.15^\circ$ & $-69.2^\circ$ &  & $60^\circ$ \\ 
             \cline{2-4} \cline{6-6} 
             & SWGO & $-24.2^\circ$ & $-69.15^\circ$ &  & $30^\circ$ 
             \\ \cline{2-4} \cline{6-6} 
             & \begin{tabular}[c]{@{}c@{}} CTA\\ (MST North)\end{tabular} & $28.45^\circ$ & $-17.53^\circ$ &  & $7^\circ$ 
             \\
             \cline{2-4} \cline{6-6} 
             & \begin{tabular}[c]{@{}c@{}} CTA\\ (SST South)\end{tabular} & $-24.41^\circ$ & $-70.24^\circ$ &  & $8^\circ$ 
             \\ \hline
        \end{tabular}
        \caption{Neutrino and gamma-ray experiments considered in this work, with their respective locations in terms of latitude and longitude. For neutrino experiments we show the corresponding volume of the detector with respect to IceCube, and for gamma-ray experiments we indicate the aperture angle of the detectors used to compute the field of view in Fig.~\ref{fig: 1hcoverage}.}
        \label{tab:Experiments_Location}
\end{table}
%

%%%%%%%%%%%%%%%
\begin{figure*}[ht]
  \centering
  \includegraphics[width=\textwidth]{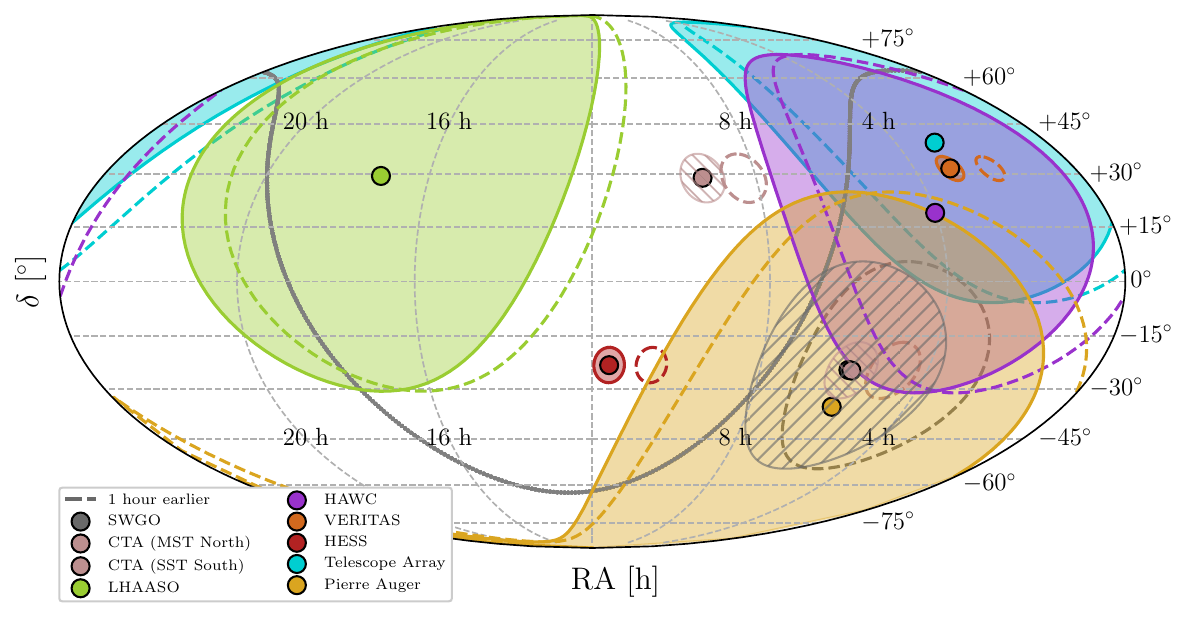}
    \caption{Illustration of the field of view in equatorial coordinates covered by existing gamma-ray experiments:  HAWC~\cite{historical:2023opo,HAWC:2019wla} (purple) and LHAASO~\cite{DiSciascio:2016rgi,LHAASO:2019qtb} (light green), VERITAS~\cite{VERITAS:2006lyc,Archambault:2017asc} (orange), HESS~\cite{HESS:2018pbp,HESS:2023zzd} (red), Telescope Array~\cite{TelescopeArray:2012qqu,TelescopeArray:2012uws} (cyan) and Pierre Auger~\cite{PierreAuger:2015eyc} (golden). We also include future experiments, represented by the hatched areas: SWGO (gray)~\cite{SWGO:2025taj,Chiavassa:2025aym} and CTA (light brown)~\cite{CTA-SST-1Mproject:2021kqa}.  The colored regions denote the instantaneous field of view of these experiments on February 13th at 01:00:00 UTC, our benchmark date. The area encompassed by the dashed line represents the instantaneous field of view of the corresponding experiment one hour earlier with respect to the event. The gray line indicates the galactic plane.}
    \label{fig: 1hcoverage}
\end{figure*}
%%%%%%%%%%%%%%%
In addition to neutrinos, high-energy photons may also play a significant role in point-like searches. Photons are much more likely to interact with gamma-ray observatories, making them key candidates for identifying point sources. However, in contrast to neutrino telescopes, gamma-ray detectors have a much more limited instantaneous field of view, which means there is a higher chance of missing a transient signal. In such cases, the short-duration burst could occur entirely outside the field of view of the gamma-ray detectors, and the network of neutrino telescopes becomes crucial for detecting the transient signal. Moreover, transient sources capable of emitting neutrinos may elude traditional electromagnetic observation entirely~\cite{Zhou:2023rtr, Murase:2019vdl, Inoue:2019yfs}, the photon signal may be delayed by years~\cite{Fang:2025nzg}, or produce only a faint sub-GeV gamma-ray signature~\cite{Murase:2015xka}, rendering them effectively invisible to conventional Earth-based gamma-ray experiments.

For the scenarios where photons accompany the neutrino emission, we show in Fig.~\ref{fig: 1hcoverage} the instantaneous field of view covered over one hour by Earth-based gamma-ray detectors HAWC~\cite{historical:2023opo,HAWC:2019wla} (purple) and LHAASO~\cite{DiSciascio:2016rgi,LHAASO:2019qtb} (light green), VERITAS~\cite{VERITAS:2006lyc,Archambault:2017asc} (orange), HESS~\cite{HESS:2018pbp,HESS:2023zzd} (red), Telescope Array~\cite{TelescopeArray:2012qqu,TelescopeArray:2012uws} (cyan), Pierre Auger~\cite{PierreAuger:2015eyc} (golden) as well as the future projects 
CTA~\cite{CTA-SST-1Mproject:2021kqa} (light brown hatched area) and SWGO\footnote{There are three possibilities for the SWGO location: Argentina (Alto Tocomar), Chile (Pampa La Bola) and Peru (Imata).   We have chosen to display the Argentinian option as it has less overlap with other gamma-ray detectors.}~\cite{SWGO:2025taj} (gray hatched area).
The location and field of view of each experiment are listed in Table~\ref{tab:Experiments_Location}. 
The field of view encompassed by the colored regions is computed at {the benchmark date
\be\label{eq:benchmark_date}
\text{Benchmark~date: February~13th~at~01:00:00~UTC},
\ee
which we use throughout the text, and the dashed lines indicate the instantaneous sky coverage one hour earlier. Notice that, while we had to choose a specific year in order to obtain the field of view depicted in Fig. \ref{fig: 1hcoverage}, say 2030, the dependence on it is very weak, amounting only to $\sim \mathcal{O}(1\%)$ in 20 years and therefore negligible for our purposes.
Still, for a given transient source, one should consider the characteristic burst duration and compute the total field of view based on it. Any event occurring outside of the total field of view, which is denoted by the colored regions in Fig.~\ref{fig: 1hcoverage}, would fail to produce a detectable photon signal, thus making neutrino telescopes the only viable option for detection.

%%%%%%%%%%%%%%%%%%%%%%%%%%%%%%%%%%%%%%%%%%%%%%%%%%%%%%%%%%%%%%%%%%%%%%%%%%%%%%%%%%%%%%%%%%%%%%%%%%%%%%%%%%%%%%%%%%%%%%%%%%%%%%%%%%%%%%%%%%%%%%%%%%%%%%%%%

\section{\label{sec:Aeff}The effective area}

The location of the source in the sky plays a central role in our analysis. One way to specify its location is through its equatorial coordinates: declination $\delta$ and right ascension (RA). The declination $\delta \in [-\pi/2, \pi/2]$ is measured relative to the celestial equator, being positive (negative) in the northern (southern) celestial hemisphere, and corresponds to a celestial latitude. The right ascension $\text{RA} \in [0, 2\pi)$ represents the angular distance along the celestial equator measured from the point where the Sun crosses it from South to North, known as the March Equinox, and is analogous to a celestial longitude. It is common practice to write RA in terms of hours, where $2\pi$ (0) corresponds to 24h (0h), such that one hour represents a change of 15$^\circ$ in RA. The parametrization in terms of equatorial coordinates $(\delta,\text{RA})$ is very useful for our purposes as it does not depend on the neutrino telescopes considered.

Because the source's position in the sky influences how neutrinos arrive at Earth and how they are detected, the event rate expected in a neutrino telescope depends on both the source coordinates and the detector response. This response is captured by the detector’s effective area, which varies with energy and the direction of arrival. The expected event rate from a transient point source located at $(\delta,\text{RA})$ is given by
\be
    \frac{\dd N_\text{evts}}{\dd t} = \frac{1}{4\pi d^2} \int \dd E \frac{\dd^2N}{\dd E \dd t} A_\text{eff}(E,\delta, \text{RA}),
    \label{eq:dNevtsdt}
\ee
where $E$ is the incoming particle energy, $\dd^2 N/\dd E\dd t$ multiplied by the factor $1/4\pi d^2$ that accounts for the isotropic propagation from a point source located at distance $d$, is the instantaneous differential flux at Earth. The effective area $A_\text{eff}$ quantifies the sensitivity of the neutrino telescope, being proportional to the neutrino cross-section, detector volume, and efficiency. Furthermore, it encodes the dependence on the source's position relative to the detector; it accounts for the attenuation of neutrinos as they traverse the Earth, which depends on the matter column density along the neutrino path to the detector~\cite{Gaisser:2016uoy}. Using Eq.~\eqref{eq:dNevtsdt}, we can perform a transient search analysis. However, the challenge is that $A_\text{eff}(E, \delta, \text{RA})$ is not publicly available, so we must engineer an effective area that accurately reflects the relative position of the source and detector at the time of the burst.

To obtain an expression for $A_\text{eff}(E, \delta, \text{RA})$, we first introduce the detector's local coordinates, $\theta$ and $\phi$. The angle $\theta \in [0, \pi]$ is the zenith angle, measured from the vertical axis at the detector, while $\phi \in [0, 2\pi]$ is the azimuthal angle. Owing to the approximate cylindrical symmetry of neutrino telescopes, the dependence on $\phi$ can be neglected. For a source located at $(\delta,\text{RA})$, we can convert to the local zenith angle as
\be
    \cos{\theta} = \sin{\lambda} \sin{\delta} + \cos{\lambda}\cos{\delta}\cos(2\pi\frac{t}{T} - \text{RA}),
\label{eq:equatorial_to_local}
\ee
where $\lambda$ is the detector latitude, $t$ is the local sidereal time, and $T$ is the duration of one sidereal day~\cite{PierreAuger:2019azx}. For the IceCube experiment, $\lambda \simeq -90^\circ$ and Eq.~\eqref{eq:equatorial_to_local} reduce to  
\be\label{eq:equatorial_to_local_IC}
    \cos\theta \simeq - \sin\delta\quad (\text{IceCube}),
\ee
eliminating any dependence on RA and time, representing an enormous simplification. As a result, IceCube can express its effective area as a function of only the neutrino energy and source declination, i.e.  $A_\text{eff}^\text{IC}(E,\delta)$. 
In contrast, for experiments not located at either geographic pole, the dependence on RA and time must be explicitly accounted for. To illustrate this, in Fig.~\ref{fig:km3NeT_vs_IC_len} we use \texttt{astropy}~\cite{Astropy:2013muo, Astropy:2018wqo, Astropy:2022ucr} to convert the equatorial coordinates into local coordinates. We show the relationship between the zenith angle $\theta$ and declination $\delta$ for IceCube (blue), KM3NeT at the corresponding selected time (solid orange), and ten hours later (dashed orange). As shown in the plot, the relationship between $\theta$ and $\delta$ varies significantly with the observation time, which in turn affects the expected number of events. To perform the change of coordinate explicitly, we have fixed in Fig.~\ref{fig:km3NeT_vs_IC_len} the benchmark date in Eq. \eqref{eq:benchmark_date}.

%%%%%%%%%%%%%%%%%%%%%%%%%%%%%%%%%%%%%%%%%%%%%%%%%%%%%%%%%%%%%%%%%
\begin{figure}[t]
    \centering
    \includegraphics[width=\linewidth]{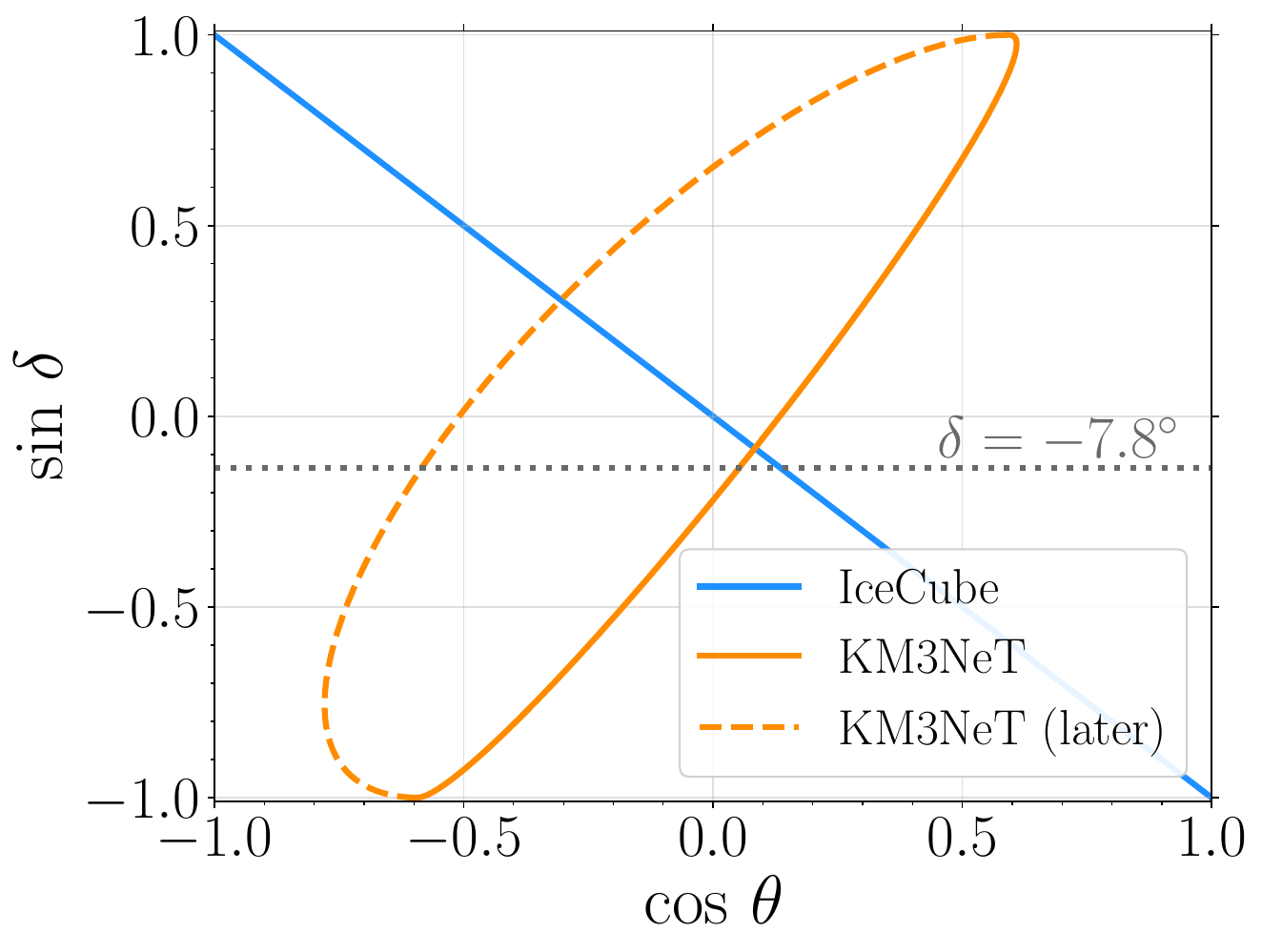}
    \caption{Relationship between source declination $\delta$ and the local zenith angle $\theta$ for IceCube (blue) and KM3NeT, assuming a fixed right ascension of $\text{RA} = 94.3^\circ$ on February 13th 01:00:00 UTC
    %13th February, 2023, at 01:16:47~UTC 
    (solid orange) and ten hours later (dashed orange). We also highlight by the dotted gray line the value corresponding to $\delta=-7.8^\circ$.}
    \label{fig:km3NeT_vs_IC_len}
\end{figure}
%%%%%%%%%%%%%%%%%%%%%%%%%%%%%%%%%%%%%%%%%%%%%%%%%%%%%%%%%%%%%%%%%
To compute the instantaneous effective areas of different neutrino experiments, we rely on the coordinate transformation discussed above. We adopt the same strategy from the Planetary Neutrino Monitoring network (PLE$\nu$M)~\cite{Schumacher:2021hhm,Schumacher:2025qca} and utilize their publicly available code~\href{https://github.com/PLEnuM-group/Plenum}{\faGithubSquare}. In this framework, all neutrino telescopes are assumed to share the same effective area as IceCube---when expressed in local coordinates---with corrections applied to account for differences in detector size (see Table~\ref{tab:Experiments_Location}, ``Volume scaling"). Specifically, they consider the IceCube effective area for muon tracks~\cite{IceCube:2021xar}, which is provided as a function of the source declination and neutrino energy, $A_\text{eff}^\text{IC}(E, \delta)$. This can be recast in terms of local coordinates using Eq.~\eqref{eq:equatorial_to_local_IC}, yielding the effective area as a function of the neutrino energy and zenith angle, $A_\text{eff}^\text{IC}(E, \theta)$.

Given the detection time, source location $(\delta, \text{RA})$, and the geographic coordinates of each experiment (listed in Table~\ref{tab:Experiments_Location}), \texttt{astropy} can be used to determine the corresponding zenith angle $\theta$ for each detector. With this information, the effective area can then be computed using $A_\text{eff}^\text{IC}(E, \theta)$.  Since the mapping from equatorial to local coordinates depends on both time and detector location, the resulting zenith angles, and therefore the values of the resulting effective area, will differ across experiments.

An additional complication is the fact that the publicly available effective area from IceCube is provided as an average over all right ascensions, which is reasonable given that IceCube’s sensitivity is approximately independent of RA and time. This averaged effective area is thus well-suited for analyses of diffuse fluxes or transient events lasting longer than a day. However, our focus is on short-lived transient events, with burst durations of  $\mathcal{O}(100~\text{s})$ for which the RA dependence is important. To restore it, we bin the IceCube effective area uniformly in RA and normalize it so that integrating over RA recovers the original result. In this manner we were able to get  $A_\text{eff}^\text{IC}(E,\delta,\rm RA)$ and in local coordinates  $A_\text{eff}^\text{IC}(E,\theta,\phi)$ and use the same strategy of PLE$\nu$M to compute the instantaneous effective areas of different detectors.
This procedure is justified by IceCube’s unique location at the South Pole. While this is similar to the PLE$\nu$M strategy for adapting IceCube’s effective area to other detectors, their approach involves integrating over RA for each detector after converting to local coordinates. In contrast, we retain the RA dependence throughout, yielding an effective area that better captures the properties of short-duration transient signals.

In the first two rows of Fig.~\ref{fig:IC_vs_KM3NeT_eff_area_and_len}, we exhibit the instantaneous effective areas for IceCube and KM3NeT on our benchmark date\,\eqref{eq:benchmark_date}. The left panels display the effective areas, while the right panels show the reconstructed distance the neutrino travels through the Earth in local coordinates for an event at $(\delta,\rm{RA}) =(-7.8^\circ, 94.3^\circ)$. 
The third row of Fig.~\ref{fig:IC_vs_KM3NeT_eff_area_and_len} presents the same analysis but for ten hours later. 
At that time, a source producing neutrinos with energies larger than 10 PeV appears at a less favorable position relative to KM3NeT, with neutrinos traversing a longer path through the Earth. This reduces the effective area and, consequently, the chance of detection.
Our results highlight that, for short-duration bursts, the relative sensitivity of neutrino detectors can change significantly with the event time and source location, underscoring the need for a coordinated global network of neutrino telescopes to effectively monitor transient events.
Note, however, that the sensitivity of neutrino observatories depends on the source’s energy spectrum, as changes in effective area are strongly energy-dependent.

In the following, we study PBHs as benchmark candidates for transient neutrino sources. Using the framework developed in this section, we compute the effective areas of the neutrino experiments listed in Table~\ref{tab:Experiments_Location}. We then explore how combining information from different detectors can enhance sensitivity to such events and improve source localization.

\begin{figure*}[t]
    \centering
        \begin{minipage}[t]{0.43\textwidth}
        \centering
        \includegraphics[width=\textwidth]{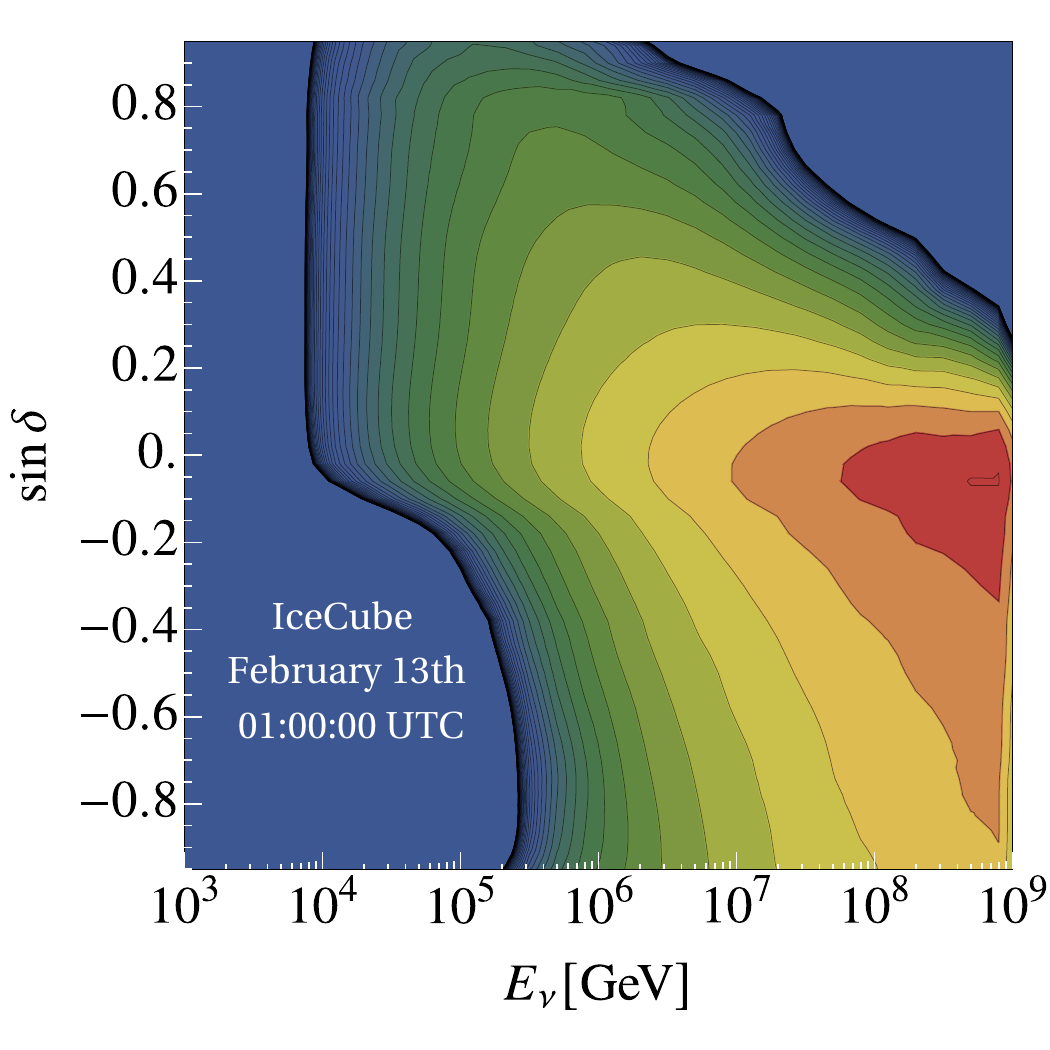}
    \end{minipage}
    \hfill
    \begin{minipage}[t]{0.4\textwidth}
        \centering
        \includegraphics[width=\textwidth]{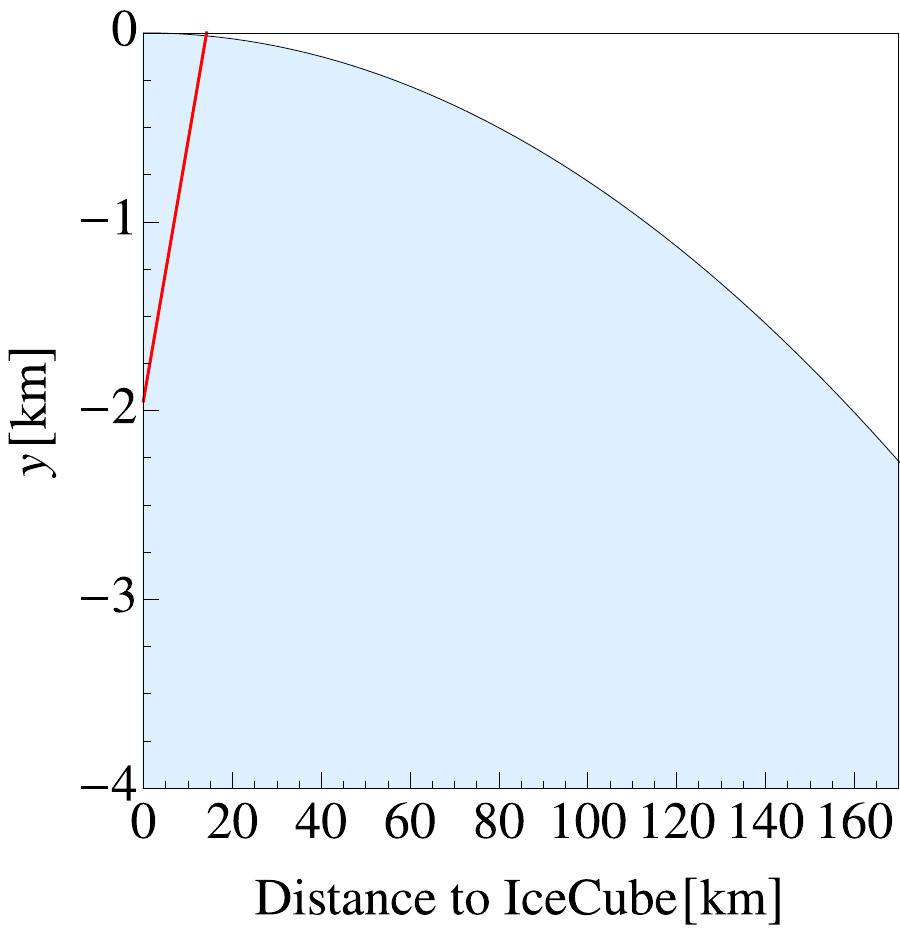}
    \end{minipage}

    \begin{minipage}[t]{0.43\textwidth}
        \centering
        \includegraphics[width=\textwidth]{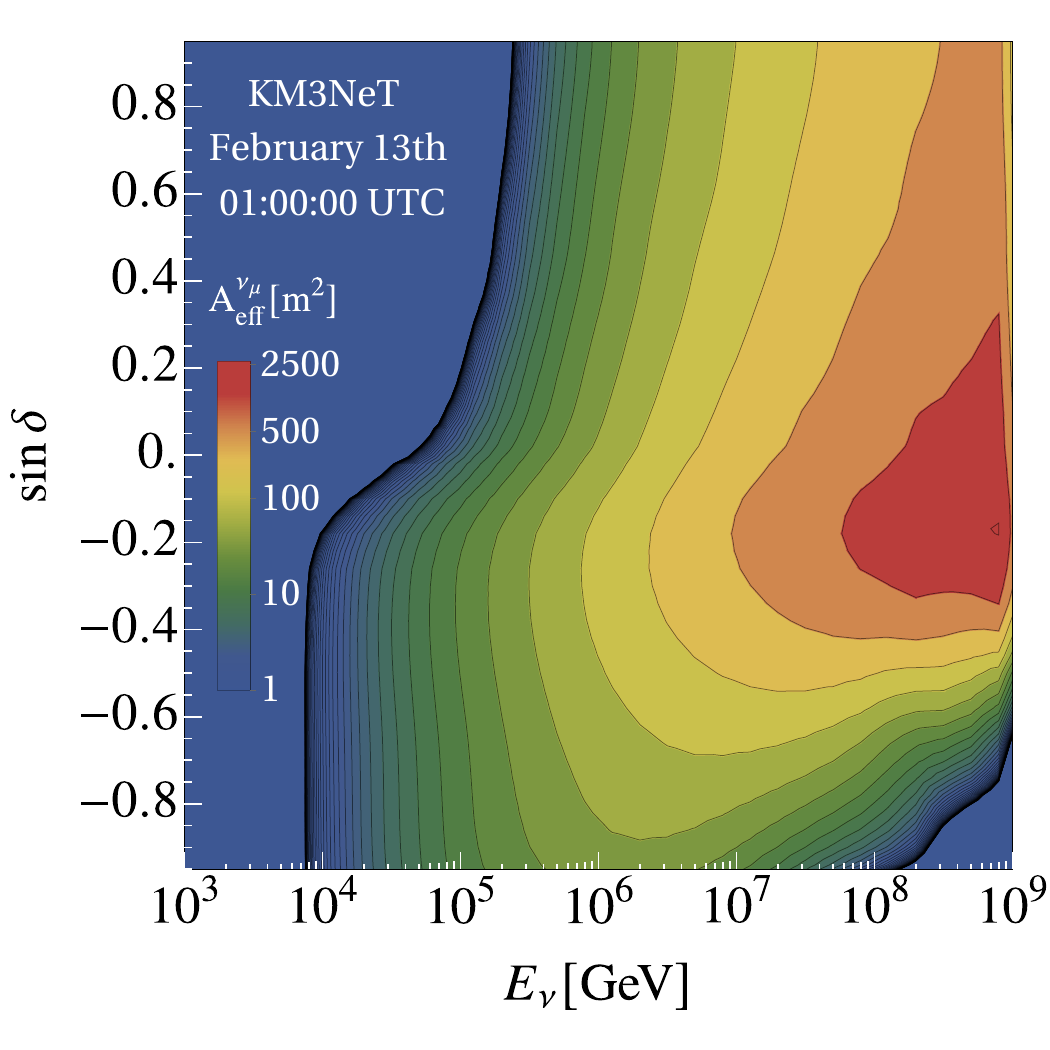}
    \end{minipage}
    \hfill
    \begin{minipage}[t]{0.41\textwidth}
        \centering
        \includegraphics[width=\textwidth]{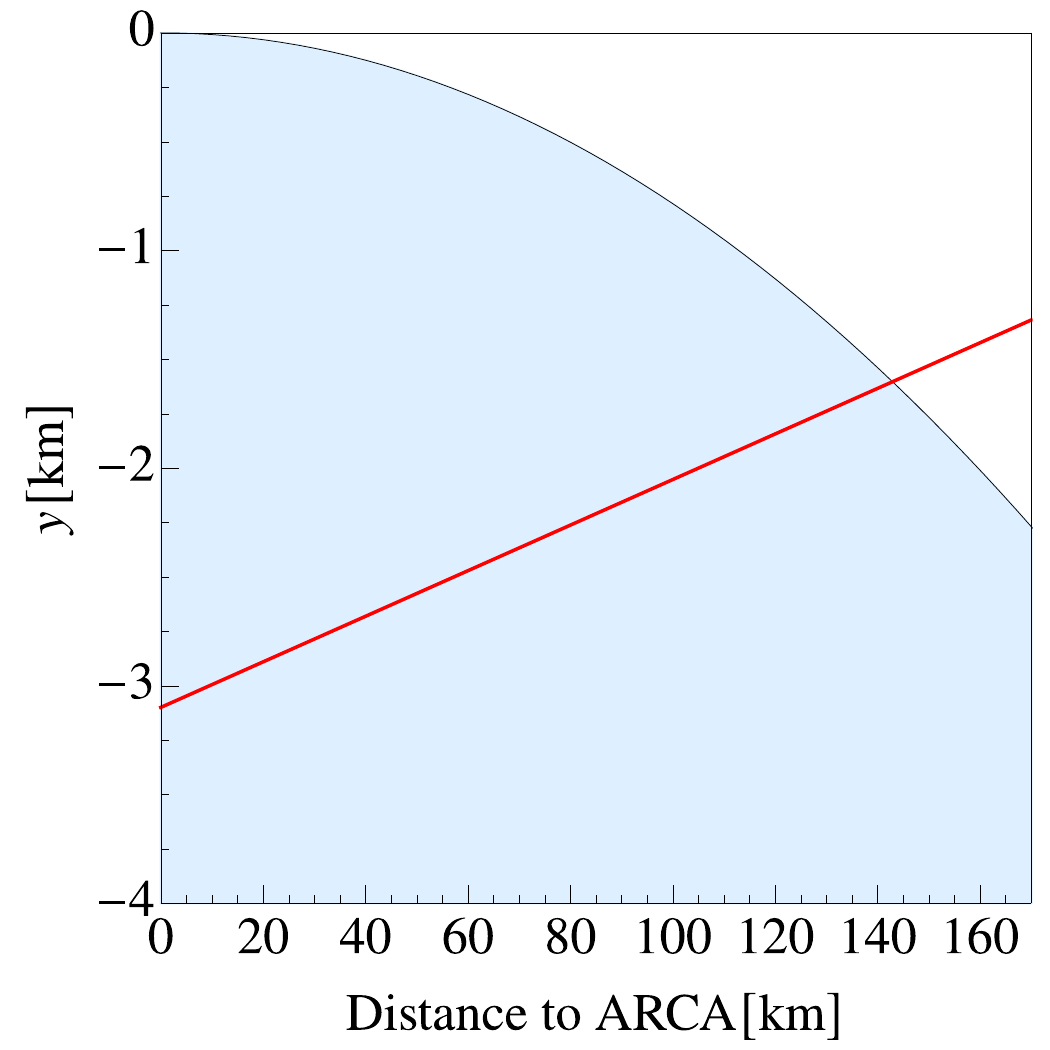}
    \end{minipage}
    
    \begin{minipage}[t]{0.43\textwidth}
        \centering
        \includegraphics[width=\textwidth]{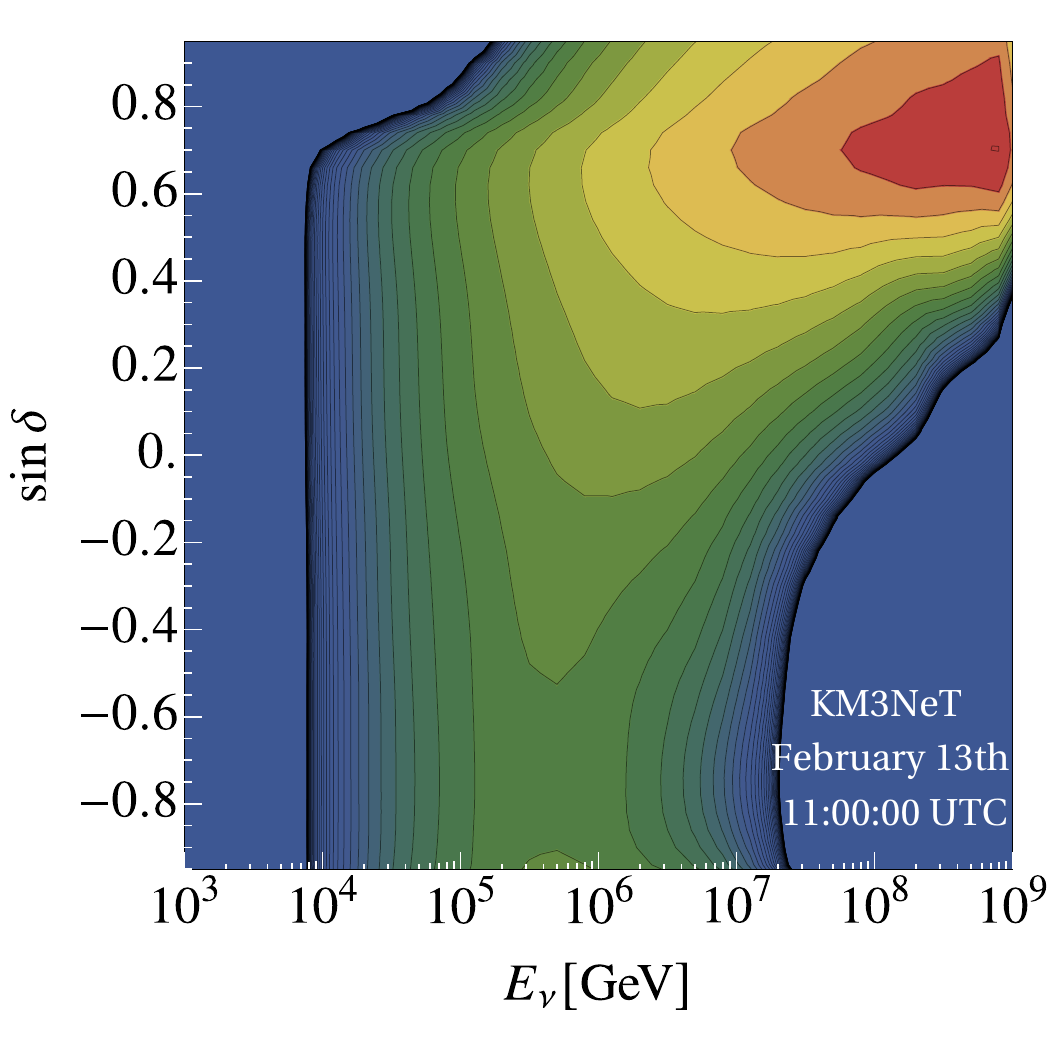}
    \end{minipage}
    \hfill
    \begin{minipage}[t]{0.41\textwidth}
        \centering
        \includegraphics[width=\textwidth]{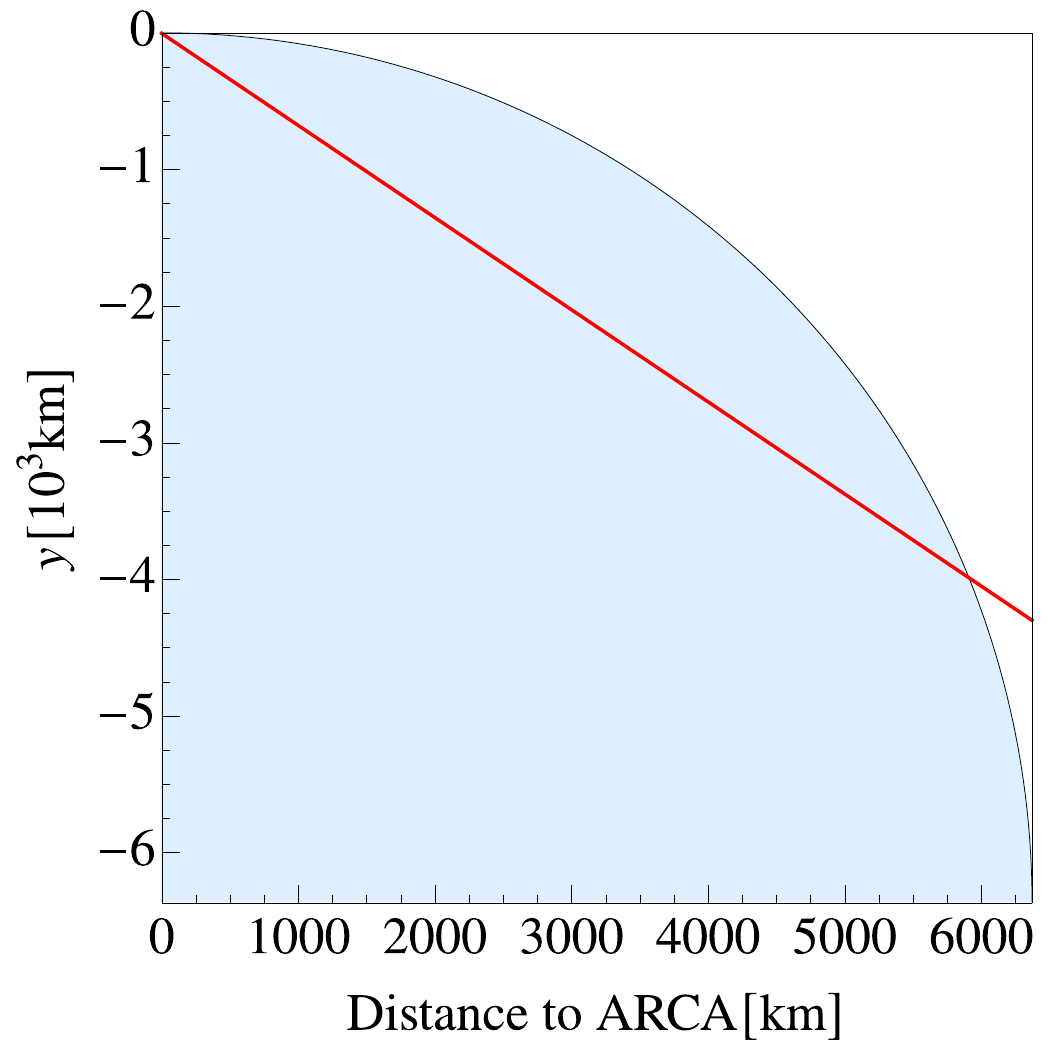}
    \end{minipage}
\vglue -0.1cm
    \caption{The top and middle left panels show the effective areas of IceCube and KM3NeT, respectively, on February 13th at 01:00:00 UTC. The bottom left panel shows the same information for KM3NeT, but evaluated ten hours later. The right panels displays the reconstructed neutrino trajectory through the Earth, shown in local coordinates for an event at $(\delta,\rm{RA}) =(-7.8^\circ, 94.3^\circ)$.}
    \label{fig:IC_vs_KM3NeT_eff_area_and_len}
\end{figure*}

\section{Case study: Bursts from a Primordial Black Hole}
\label{sec:PBHs}

Primordial Black Holes are hypothetical black holes that could have formed in the early Universe, possibly within the first fractions of a second after the Big Bang, through different production mechanisms~\cite{Carr:2009jm, Carr:2020gox, Flores:2024eyy}. Unlike black holes formed via stellar collapse, PBHs are not restricted to a narrow mass range and could span a wide spectrum, from the Planck mass up to many solar masses.

\subsection{Hawking evaporation}\label{sec:Hawking}

Once formed, PBHs are expected to evaporate via Hawking radiation~\cite{Hawking:1974rv,Hawking:1975vcx,Page:1976df}, a semi-classical process in which quantum fluctuations in a classical curved spacetime lead to the emission of particles. The expected number of particles emitted per unit time and energy by a Schwarzschild black hole\footnote{Although black holes are characterized by their mass, spin and charge, a PBH typically loses charge and spin faster than mass in the standard scenario. Thus, by in the end of their evaporation, PBHs are well approximated by Schwarzschild black holes~\cite{Page:1976df,PhysRevD.14.3260,PhysRevD.16.2402}.} (i.e., a non-rotating, uncharged black hole) is given by
\begin{align}
\label{eq: SpectrumEmittedParticles}
    \frac{\dd^2 N_j}{\dd E \dd t} = \frac{g_j}{2\pi}\frac{\Gamma_{s_j}(E)}{\exp(8\pi G  M  E) - (-1)^{2s_j}}\, ,
\end{align}
where $G$ is Newton's constant, $M$ is the mass of the PBH at time $t$, $j$ denotes the species of the emitted particle with spin $s_j$, energy $E$ and internal number of degrees of freedom $g_j$. The factor $\Gamma_{s_j}$ is the so-called gray-body factor, which accounts for deviations from a perfect black-body spectrum due to the gravitational potential of the black hole~\cite{Hawking:1975vcx,Hawking:1974rv,Unruh:1976fm,Doran:2005vm, Lunardini:2019zob,Masina:2021zpu,Cheek:2021odj}. This resemblance to black-body radiation further allows one to associate a temperature to the black hole, given by
\begin{align}
\label{eq: TempPBH}
     T = \frac{1}{8\pi G M} \sim 1~{\rm TeV} \left(\frac{10^{10}~{\rm g}}{M}\right),
\end{align}
such that it increases for lighter black holes. 

The evaporation process is derived in a semi-classical framework, where quantum fields evolve on a fixed classical curved spacetime. This approximation is expected to hold until the black hole approaches the Planck scale, beyond which quantum gravity effects become significant and the final fate of the PBH remains uncertain\footnote{
Even before this point, the information loss problem introduces further ambiguity: after the \emph{Page time}~\cite{Page:1993wv,Page:2013dx,Perez-Gonzalez:2025try}, the von Neumann entropy of the Hawking radiation exceeds the thermodynamic Bekenstein-Hawking entropy, suggesting that non-thermal corrections to the spectrum may arise. Since the PBHs considered here would have passed the Page time long ago, such effects could in principle alter the spectrum~\cite{Perez-Gonzalez:2025try}. However, given the uncertainty on this problem~\cite{Buoninfante:2021ijy,Buoninfante:2025gqk}, we assume the validity of the semi-classical approximation up to near-Planckian scales.}. Hereafter, full evaporation refers to the black hole reaching the Planck mass, beyond which its fate remains unknown.

As particles are produced through Hawking radiation, energy conservation implies that the black hole loses mass. The mass loss rate is given by~\cite{MacGibbon:1990zk,MacGibbon:1991tj,Cheek:2021odj}
\be\label{eq:dMdt}
    \frac{\dd M}{\dd t} = - \sum_j\int \dd E~E \frac{\dd^2 N_j}{\dd E \dd t} \simeq -\frac{4\times 10^{-3}}{G^2 M^2},
\ee
where we assumed that only Standard Model particles are emitted and neglected particle masses in the emission. This last approximation is only valid for $M\lesssim 10^{12}~{\rm g}$, above which some particle masses become larger than the PBH temperature, making their emission Boltzmann suppressed and reducing the value of the constant. Under these assumptions, from Eq.~\eqref{eq:dMdt}, we can estimate the time for the black hole to fully evaporate, i.e., its lifetime
\begin{equation}
    \tau \sim \left(\frac{M_0}{10^{15}~\text{g}}\right)^3\times10^{18}~\text{s},
    \label{eq:lifetime}
\end{equation}
where $M_0$ is black hole initial mass. A PBH with $M_0\sim 10^{15}~\text{g}$ would be completing its evaporation today and could potentially be detected by neutrino and gamma-ray telescopes.

The cubic scaling of $\tau$ with the PBH mass implies that a small decrease in mass drastically shortens the PBH lifetime. Combined with the rising temperature, it leads to a sharp, explosive finale, the so-called final burst, when the PBH radiates intensely within a short period. This final explosion defines PBHs as transient sources, making them suitable targets for the framework developed in Section~\ref{sec:Aeff}.

It is also useful to write the PBH temperature at $\tau_B$ seconds before complete evaporation. By combining Eq.~\eqref{eq: TempPBH} and Eq.~\eqref{eq:lifetime}, we obtain
\begin{align}
\label{eq:TempTime}
    T\simeq 7.8\times 10^3 \times\left(\frac{1~\text{s}}{\tau_B} \right)^{1/3}~\text{GeV},
\end{align}
indicating that for $\tau_B \sim \mathcal{O}(100~\text{s})$, the PBH predominantly emits particles with energies above the TeV scale.

We are interested in the spectrum of photons and neutrinos produced during the PBH evaporation. These particles can be emitted in two ways: directly from the PBH itself by Hawking radiation, known as the primary spectrum, or from the decays of primary particles, generating a secondary spectrum. We will also assume that the primary neutrino emission produces neutrinos in the mass basis. To compute these spectra, we use the publicly available code \texttt{BlackHawk}~\cite{Arbey:2019mbc,Arbey:2021mbl}, with the hadronization handled by \texttt{HDMSpectra}~\cite{Bauer:2020jay}. For more details on the spectrum and burst rate, we refer the reader to App.~\ref{app:burst}.

\subsection{Number of events}

In what follows, we consider a single PBH evaporating within our Solar System, located at a distance $d_{\text{PBH}}$ from Earth and positioned at equatorial coordinates $(\delta_\text{PBH},\text{RA}_\text{PBH})$. We assume this PBH originates from a broader population whose mass and spatial distributions are consistent with current observational bounds on PBH burst rates~\cite{Archambault:2017asc,Fermi-LAT:2018pfs,Abdo:2014apa,HESS:2021rto,HAWC:2019wla}. Within these bounds, the existence of a single nearby PBH remains allowed~\cite{Boluna:2023jlo,DeRomeri:2024zqs}. 

The expected differential event rate at Earth of neutrinos with flavor $\alpha$ for the primary spectrum is
\al{\label{eq:dNevtdt_nu_primary}
    \frac{\dd N_\text{evt}^{\nu_\alpha}}{\dd t}\Big|_\text{pri} 
    & = \frac{1}{4\pi d_\text{PBH}^2}\sum_{i=1,2,3} |U_{\alpha i}|^2\\
    & \times \int \dd E  \frac{\dd^2 N_{\nu_i}}{\dd E \dd t}\Big|_\text{pri} A_\text{eff}(E,\delta_\text{PBH},\text{RA}_\text{PBH}),
}
where $\dd^2 N_{\nu_i}/\dd E\dd t|_\text{pri}$ is the spectrum for the mass eigenstate $i$~\cite{Lunardini:2019zob}. The elements of the Pontecorvo--Maki--Nakagawa--Sakata (PMNS) mixing  matrix $U$ are included to properly account for neutrino propagation and flavor oscillations from the emission point to the Earth.
For the secondary neutrino spectrum, the differential $\nu_\alpha$ event rate at Earth 
is
\al{\label{eq:dNevtdt_nu_secondary}
    \frac{\dd N_\text{evt}^{\nu_\alpha}}{\dd t}\Big|_\text{sec}  
    & = \frac{1}{4\pi d_\text{PBH}^2}\sum_{\beta=e,\mu,\tau}\sum_{i=1,2,3} |U_{\alpha i}|^2|U_{\beta i}|^2\\
    & \times \int \dd E  \frac{\dd^2 N_{\nu_\beta}}{\dd E \dd t}\Big|_\text{sec} A_\text{eff}(E,\delta_\text{PBH},\text{RA}_\text{PBH}),
}
with $\dd^2 N_{\nu_\beta}/\dd E \dd t|_\text{sec}$ the secondary neutrino spectrum of flavor $\beta$.
For photons, the corresponding expected number of events is given by expressions analogous to Eq.~\eqref{eq:dNevtdt_nu_primary} and Eq.~\eqref{eq:dNevtdt_nu_secondary}, but without the PMNS matrix factors. The total number of neutrino events of flavor $\nu_\alpha$ detected during the final $\tau_B$ seconds before complete evaporation is obtained by integrating the differential event rates over a time interval
%m
\be\label{eq:Nevt_PBH}
    N_\text{evt}^{\nu_\alpha}(\tau_B)\equiv \int_{\tau-\tau_B}^{\tau}\dd t \frac{\dd N_\text{evt}^{\nu_\alpha}}{\dd t},
\ee
where $\dd N_\text{evt}^{\nu_\alpha}/\dd t$ is the sum of primary and secondary spectra in Eq.~\eqref{eq:dNevtdt_nu_primary} and Eq.~\eqref{eq:dNevtdt_nu_secondary}. Similarly, for photons, the number of events is computed using Eq.~\eqref{eq:Nevt_PBH} with the replacement $N_\text{evt}^{\nu_\alpha} \to N_\text{evt}^{\gamma}$.

\subsection{Results}

To highlight scenarios where neutrino telescopes are the only viable option for detecting a transient signal, we must first examine when gamma-ray detectors could probe a PBH final burst. From Eq.~\eqref{eq:TempTime}, we infer that detectors with energy thresholds of a few TeV are primarily sensitive to the final few hundred seconds of the burst, while detectors with energy thresholds around tens of GeV could probe not only the final burst itself, but also the diffuse emission of particles over timescales of several years, depending on the PBH's distance. We anticipate that $d^\text{max}_\text{PBH}\sim 10^{-3}~\text{pc}$ is the typical maximum distance neutrino telescopes are sensitive to~\cite{DeRomeri:2024zqs,Capanema:2021hnm,Perez-Gonzalez:2023uoi}, which should be compared to the reach of the gamma-ray detectors, $d^{\text{max}}_\text{PBH}\sim 10^{-1}~\text{pc}$~\cite{HAWC:2019wla, Yang:2024vij}. This implies that, for PBHs close enough to be detected by neutrino experiments, a gamma-ray detector would fail to observe the signal only if the source remains entirely outside its field of view during the entire time that the PBH flux exceeds the detector’s energy threshold.

As a concrete example, we again consider the same date and time of the benchmark \eqref{eq:benchmark_date} we chose previously.
The candidate gamma-ray experiments with larger sky coverage and with energy thresholds suitable for a burst analysis which are currently operating are HAWC, LHAASO, Telescope Array, and Pierre Auger as well as the future projects CTA and SWGO are shown in Fig.~\ref{fig: 1hcoverage}. If a hypothetical PBH  event at that date falls only within the field of view of Pierre Auger (outside of the field of view of other gamma-ray detectors), 
%however in Pierre Auger, 
due to its high energy threshold, $E \gtrsim 3 \times 10^{9}$~GeV, the PBH flux is significantly suppressed. Assuming an effective area of 3000~km$^2$ and 100~\% detection efficiency, which is a good approximation for particles with energies above the threshold~\cite{PierreAuger:2010zof}, we estimate the expected number of events to be $N_\text{evt} \simeq 0.5\times(10^{-4}~\text{pc}/d_\text{PBH})^2$, such that the possibility that Pierre Auger could miss the transient signal is not excluded. 

Next, we must investigate whether the gamma-ray experiments could observe an early signal, serving as pre-burst monitors.
From Ref.~\cite{Yang:2024vij}, we note that the effective area for the water Cherenkov detector array (WCDA) at LHAASO extends to the GeV range, enabling it to detect photons from a PBH well before the final explosion. We find that for $d_\text{PBH}\sim \mathcal{O}\left(10^{-4}~\text{pc}\right)$, the number of photon events expected from the PBH exceeds the background by a large margin, even a few months before evaporation. Therefore, even if LHAASO were to miss the final burst itself, having the PBH within its field of view a day earlier would be sufficient for detection. A similar argument applies to other gamma-ray detectors with comparable energy thresholds, such as HAWC or the future SWGO. Pierre Auger does not contribute to this analysis, as it is sensitive only to the final seconds of the burst, detecting photons exclusively above the EeV scale. This discussion is of particular relevance to interpreting, for instance, the KM3-230213A event as a PBH, and we leave a dedicated analysis for our companion paper~\cite{Airoldi:2025opo}.

Therefore, to examine the potential for capturing the final burst  through neutrino experiments -- while simultaneously preventing gamma-ray detectors from dominating the detection landscape, we highlight in the top panel of Fig.~\ref{fig:Nevts_PBH} the region of the sky  visible to gamma-ray observatories, shown as a gray area. This includes the full daily sky coverage of HAWC and LHAASO, as well as the instantaneous field of view of the Pierre Auger Observatory, as a dashed gray area, on our benchmark date\,\eqref{eq:benchmark_date}\footnote{We did not explicitly include the coverage of SWGO as the site is still under discussion, but it will certainly fall within the area covered by Pierre Auger.}. To make this point clearer, we recall that Pierre Auger is only sensitive to the final burst, meaning that its field of view matters only at the time of the burst itself, not before. As a result, the white region, representing parts of the sky accessible exclusively to neutrino detectors, also varies with time, being always complementary to the instantaneous coverage of Auger. We exclude the Galactic plane, as the intense background from this region makes it significantly more challenging for the gamma-ray experiments to identify transient signals. 
The colored points are obtained by transforming the terrestrial coordinates of the neutrino telescopes 
in Table~\ref{tab:Experiments_Location} to celestial ones.
From the figure, we observe that gamma-ray detectors cover a substantial fraction of the sky and would typically take the lead in the PBH search if it lies within the gray area. Nonetheless, for PBHs located outside the gray region, only neutrino telescopes can provide viable coverage, making their synergy essential for detection. 
%%%
\begin{figure*}[t]
    \centering
    \includegraphics[width=0.8\textwidth]{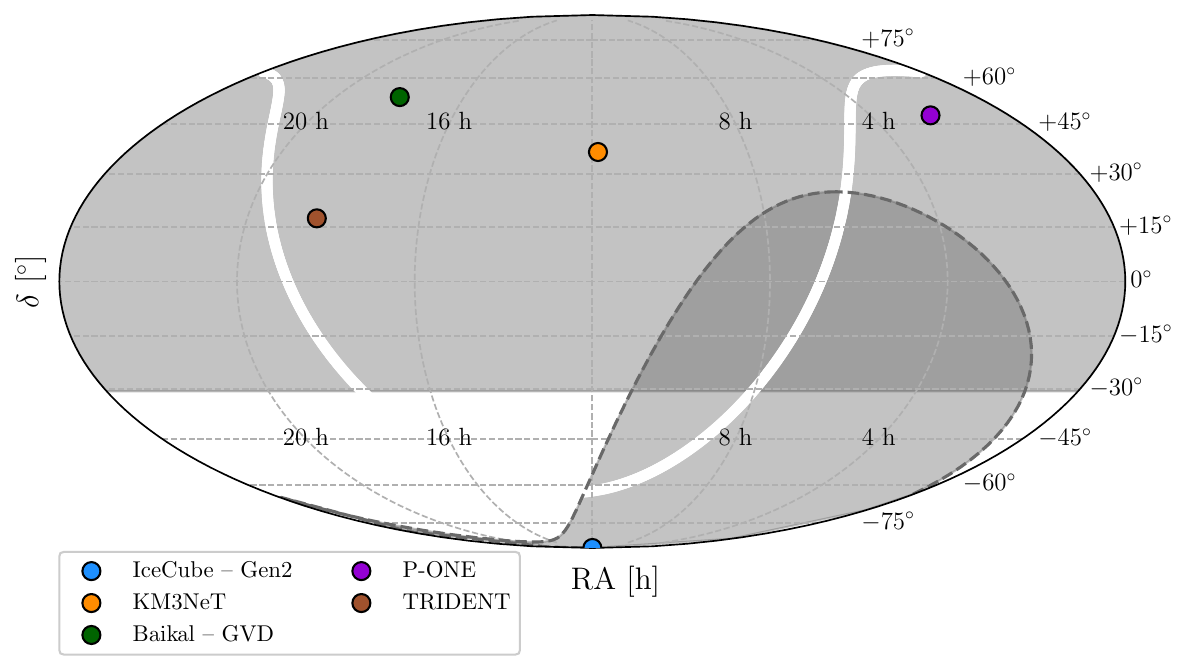}
    \centering
    \includegraphics[width=0.8\textwidth]{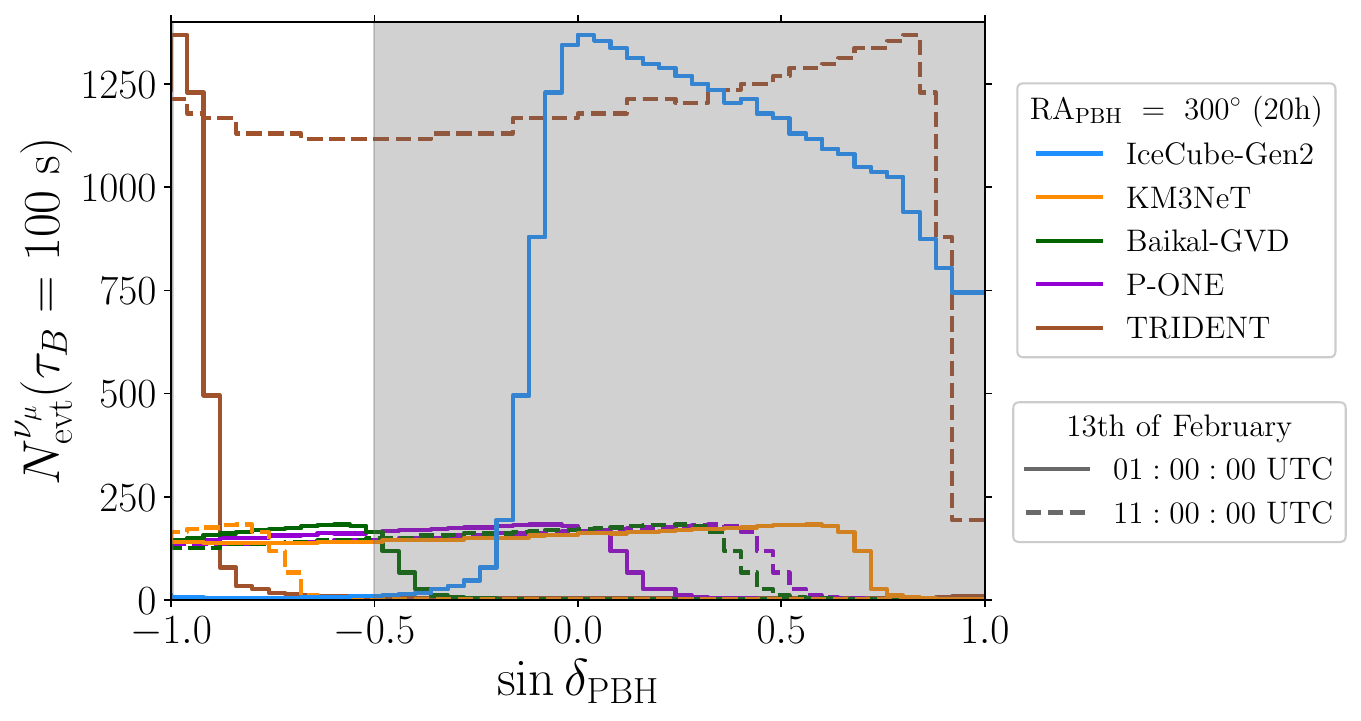}
\caption{\textbf{Top:} Portion of the sky in equatorial coordinates that is covered by gamma-ray detectors. The gray region denotes the region where a PBH could have been observed by gamma-ray experiments. The upper gray region is the one-day coverage by LHAASO and HAWC, while the region with the dashed contour is the instantaneous field of view of Pierre Auger for February 13th 01:00:00 UTC. The colored points indicate the location 
in the sky obtained by transforming the terrestrial coordinates the neutrino telescopes.}
    \label{fig:Nevts_PBH}
\end{figure*}
%%%

Assuming a PBH  located outside the field of view of gamma-ray detectors, we now turn to a detailed analysis of the joint capabilities of various present and future neutrino telescopes. 
IceCube dedicated searches for track-like muon events have an energy threshold around 100~GeV and could, in principle, probe the neutrino flux over an extended time window. However, as the region we are interested in lies above the horizon, the irreducible background from atmospheric neutrinos and muons would dominate over most of the PBH signal. Nevertheless, the steep rise in emission during the final few hundred seconds of a PBH’s life results in a sharp, background-free neutrino signal, offering a clean observational window. KM3NeT/ARCA, optimized for neutrinos above the TeV scale, is similarly sensitive only to this final burst phase. To enable a joint analysis across experiments, we therefore restrict our focus to the last stages of PBH evaporation.

In the lower panel of Fig.~\ref{fig:Nevts_PBH}, we show the number of muon neutrino events expected during the last $\tau_B = 100$~s of the PBH burst at a distance $d_\text{PBH}=10^{-4}~\text{pc}$, as a function of the source declination. We include the results for IceCube-Gen2 (blue), KM3NeT (orange), Baikal-GVD (green), P-ONE (purple), and TRIDENT (brown). Solid lines represent the number of events evaluated on the benchmark date\,\eqref{eq:benchmark_date}, while dashed lines correspond to the hypothetical scenario of the event happening ten hours later. We fix the source right ascension to $\text{RA}_\text{PBH}=300^\circ$ (20~h) and highlight the declination interval $\sin(\delta_{\text{PBH}}) \in [-1.0, -0.50]$ for which the signal would be invisible to gamma-ray experiments. Due to their different locations around the globe, each experiment is sensitive to the PBH burst in distinct ranges of declination, varying also considerably with time. In particular, we see that IceCube has no sensitivity in the allowed range $\sin(\delta_{\text{PBH}}) \in [-1.0, -0.50]$, as its effective area is suppressed at these declinations. In contrast, the other neutrino telescopes, all located in the northern hemisphere, can detect a substantial number of events. KM3NeT, P-ONE, and Baikal-GVD exhibit similar responses due to their comparable sizes, whereas TRIDENT is expected to collect significantly more events owing to its larger detection volume (see Table~\ref{tab:Experiments_Location}). This conclusion remains valid at different times, since changing the observation time effectively shifts right ascension by  $15^\circ$ per hour, and IceCube's effective area is independent of RA. Among the detectors capable of observing events, KM3NeT and Baikal-GVD are currently operational. This demonstrates that (i) neutrino telescopes provide complementary information on the PBH, (ii) a careful treatment of the time and localization dependence of the effective areas is key for the joint analysis, and (iii) that data from neutrino experiments other than IceCube will be crucial in constraining PBHs.

To conclude, a couple of comments are in order.
First regarding the maximum distance neutrino telescopes that are similar to IceCube (same volume, see Tab.~\ref{tab:Experiments_Location}) can probe if they measure a PBH burst. For a background-free search, we adopt the Poisson upper limit of $N_{\nu_\mu} < N_{\text{max}} = 2.3$ events at 90\%~C.L.~\cite{Capanema:2021hnm}. Inverting this inequality, we find $d^{\text{max}}_{\text{PBH}} = \sqrt{N_{\rm evt}^{\nu_\mu} (\tau_B)/(4\pi N_{\rm max})} \sim 10^{-3}$~pc\footnote{This differs from the result from our companion paper~\cite{Airoldi:2025opo} as for explaining the KM3NeT event we considered realistic detector conditions, that is, we corrected the effective area to account for the fact that KM3NeT was not fully operational at the time of the event. Lowering the effective area requires the PBH to be closer to the Earth.}.

Second, more generally, we may ask how realistic it is for a PBH to lie within the Solar System. Current limits on the rate of evaporating PBHs are at the level of $< 181~{\rm pc^{-3}~yr^{-1}}$ at 99\% confidence level for a burst duration of 20~s~\cite{LHAASO:2025wgl}. Translating this into a local PBH density requires assuming a specific initial mass distribution compatible with existing observational constraints, such as those from diffuse gamma rays or cosmic rays~\cite{Boluna:2023jlo}. For a nearly monochromatic distribution, i.e., all PBHs having approximately the same mass, current bounds allow for about one evaporating PBH per year within a volume of radius one parsec~\cite{DeRomeri:2024zqs}. Other mass distributions could permit a higher local density of evaporating PBHs. Therefore, the discovery of a nearby PBH would also provide valuable information about the initial PBH population, potentially challenging or refining scenarios in which PBHs constitute a significant fraction of Dark Matter.

\section{\label{sec:conclusions} Conclusions}

In this paper, we examined in detail how current and future neutrino telescopes can contribute to the search for point-like transient sources. By keeping track of the localization and time of the source. We showed that experiments located in different parts of the Earth can provide complementary information. In the case that gamma-rays are also emitted, we discussed the coverage of present Earth-based gamma-ray detectors and the conditions under which they would be sensitive to the transient source, thus allowing for a coordinated multimessenger effort. We also emphasized the importance of neutrino observatories in regions where gamma-ray experiments are blind, making neutrinos our only window into the event. Using PBHs as a case study, we explored scenarios where the source emits both photons and neutrinos. For such cases, we computed the expected number of neutrino events at various detectors. For sources that emit only neutrinos, our results can be straightforwardly extended to any region in the sky.

%%%%%%%%%%%%%%%%%%%%%%%%%%%%%%%%%%%%%%%%%%%%%%%%%%%%%%%%%%%%%%%%%%%%%%%%%%%%%%%%%%%%%%%%%%%%%%%%%%%%%%%%%%%%%%%%%%%%%%%%%%%%%%%%%%%%%%%%%%%%%%%%%%%%%%%%%%%%%%%%%%%%%
\textcolor{white}{aaa}

\begin{acknowledgments}
We thank Pedro Machado, Matheus Hostert, 
Lisa Schumacher, Edivaldo Moura Santos, Daniel Naredo, and Miguel A. Sánchez-Conde for discussions.
G.F.S.A., L.F.T.A, and G.M.S. received full financial support from the São Paulo Research Foundation (FAPESP) through the following contracts No. 2022/10894-8 and No. 2020/08096-0, 2025/07427-7,  2020/14713-2, and 2022/07360-1, respectively. R.Z.F. is partially supported by FAPESP under contract No. 2019/04837-9, and by  Conselho Nacional de Desenvolvimento Científico e Tecnológico (CNPq). Y.F.P.G. was supported by the Consolidaci\'on Investigadora grant CNS2023-144536 from the Spanish Ministerio de Ciencia e Innovaci\'on (MCIN) and by the Spanish Research Agency (Agencia Estatal de Investigaci\'on) through the grant IFT Centro de Excelencia Severo Ochoa No CEX2020-001007-S. 

\end{acknowledgments}

\appendix

\section{PBH final burst and event rate}
\label{app:burst}

This appendix expands on two key aspects of PBH evaporation mentioned in the main text: (i) the energy spectrum of particles emitted during the final stages of PBH evaporation, offering additional insights into the temporal and spectral characteristics of the final burst; and (ii) the expected event rates in neutrino and gamma-ray detectors, with a focus on illustrating the time windows during which the signal would be free of background.

\begin{figure}[t]
    \centering
    \includegraphics[width=0.9\linewidth]{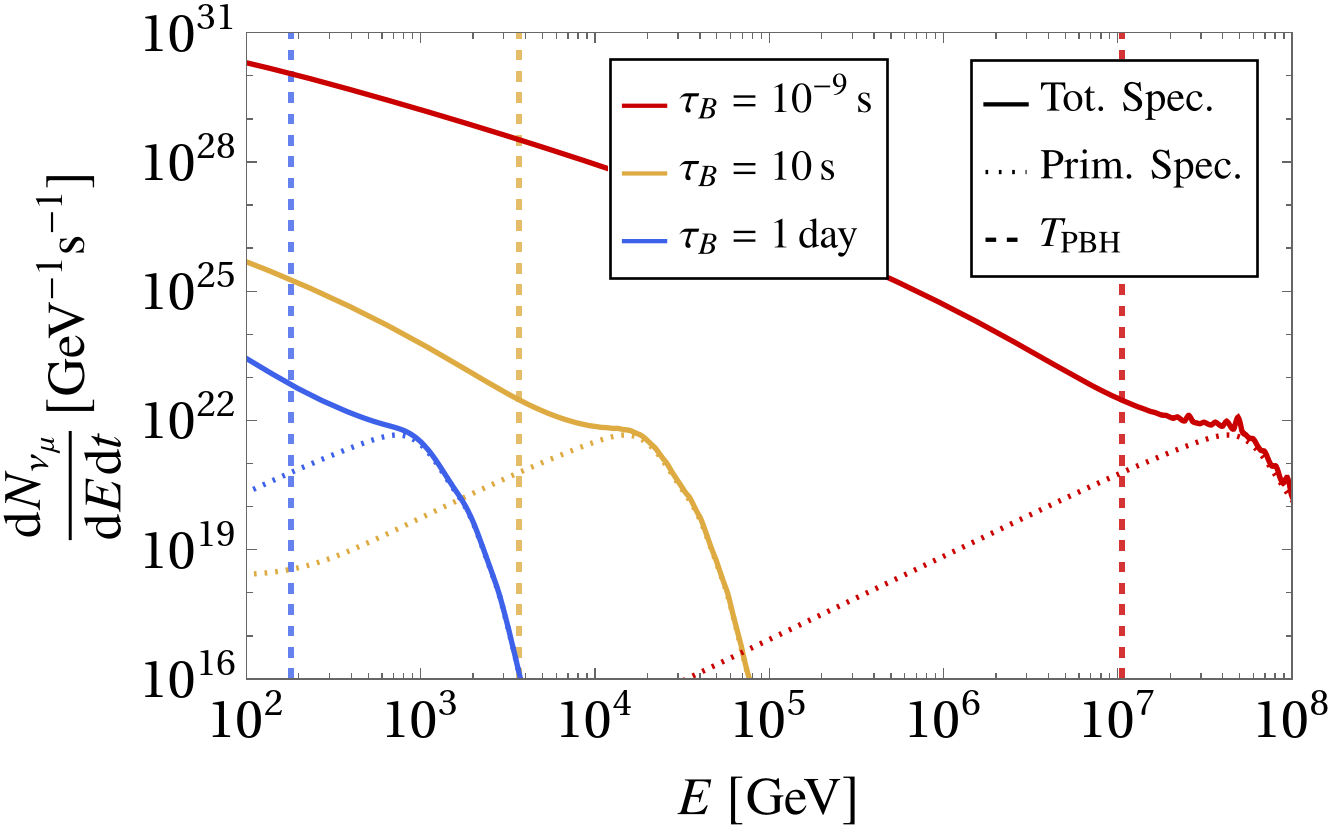}
    \caption{The PBH instantaneous differential muon neutrino spectrum $d^2N_{\nu_\mu}/dEdt$ as a function of the energy. We show curves for $\tau_B = $ 1~day (blue), 10~s (yellow) and $10^{-9}$~s (red) before the PBH fully explodes. The dashed lines indicate the PBH temperature $T_{\text{PBH}}$ at the time $\tau_B$.}
    \label{fig: InstSpec}
\end{figure}

\begin{figure}[t]
    \centering
    \includegraphics[width=0.9\linewidth]{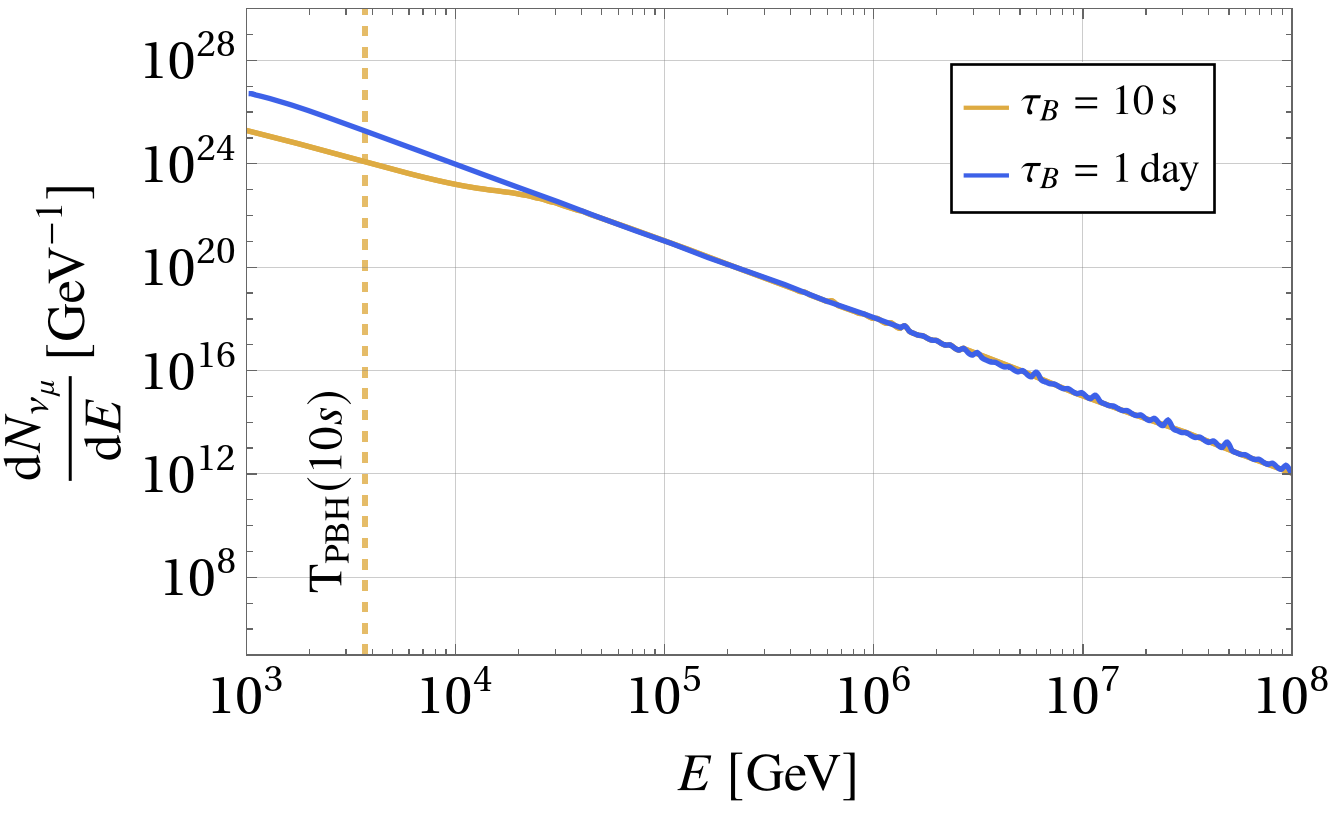}
    \caption{The time integrated Hawking spectrum for muon neutrinos $dN_{\nu_\mu}/dE$ as a function of the energy. The dashed line indicates the PBH temperature at $\tau_B = 10~\text{s}$. We note that integrating the spectrum from 1~day or 10~s before has little difference. }
    \label{fig: 10s1day}
\end{figure}

\begin{figure*}[t]
    \centering
    \begin{minipage}[h]{0.48\textwidth}
        \centering        \includegraphics[width=\textwidth]{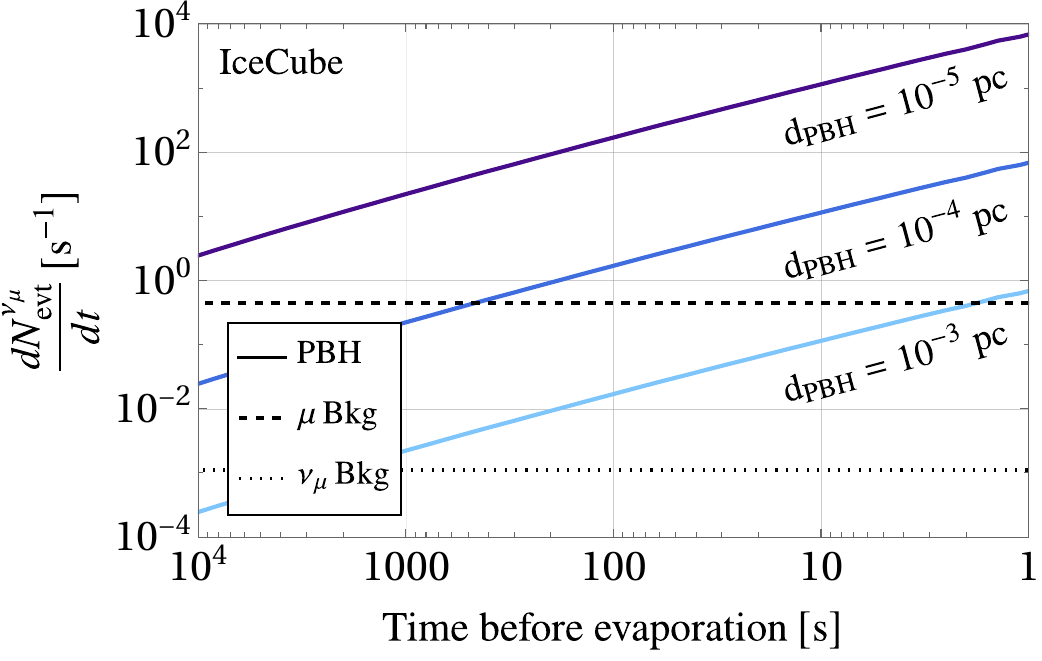}
    \end{minipage}
    \hfill
    \begin{minipage}[h]{0.48\textwidth}
        \centering        \includegraphics[width=\textwidth]{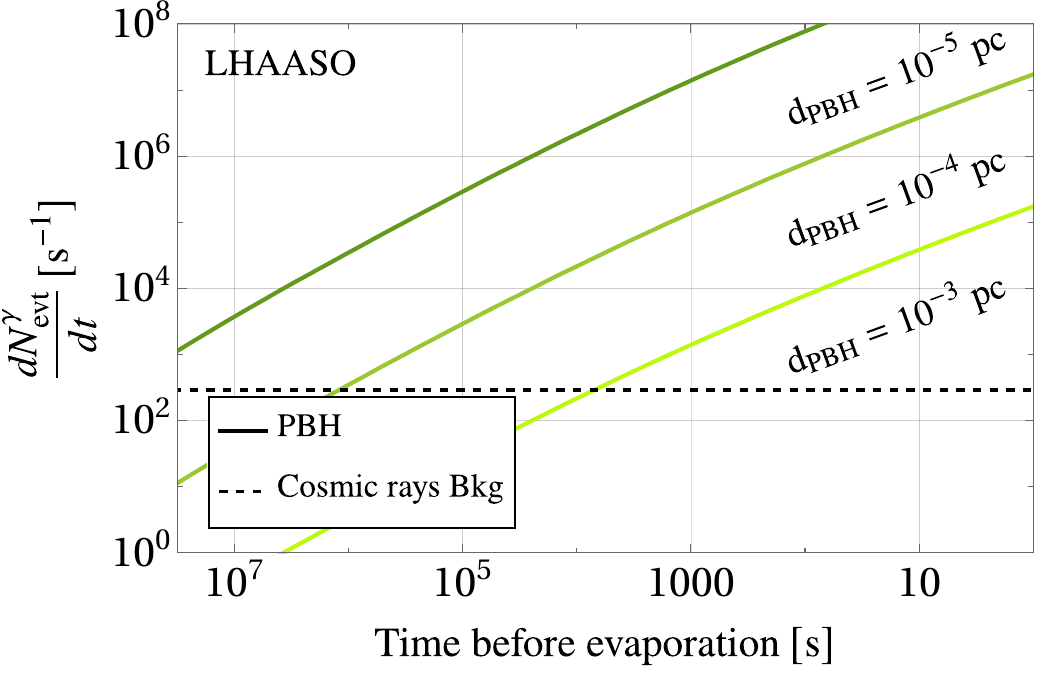}
    \end{minipage}
    \caption{\textbf{Left}: Expected differential event rate $dN_{\rm evt}^{\nu_\mu}/dt$ of track-like muon neutrinos in IceCube for a PBH located at the declination $\delta = -0.02^\circ$. The dashed line represents the muon background~\cite{IceCube:2015wro}, while the dotted line denotes the atmospheric muon neutrino background~\cite{Vitagliano:2019yzm}.
    \textbf{Right}: Expected differential photon rate $dN_{\rm evt}^{\gamma}/dt$ in LHAASO for a PBH located at $\delta = -0.02^\circ$. The dashed line represents the background from cosmic rays~\cite{Yang:2024vij,Abdo:2014apa}.
    }
    \label{fig: NevtsRate}
\end{figure*}
%%%%%%%%%%%%%

To begin, we examine the evolution of the neutrino energy spectrum as the PBH approaches its final moments. Fig.~\ref{fig: InstSpec} shows the instantaneous primary and total neutrino spectra emitted $1$~day (blue), $10$~s (yellow), and $10^{-9}$~s (red) before the PBH fully evaporates. The rapid increase in the emission rate as the remaining time shortens is accompanied by a shift of the spectrum peak towards higher energies. The dashed lines indicate the PBH temperature at the time $\tau_B$. We note that, as the PBH spectrum is not a perfect black body, the peak of the primary spectrum is not perfectly aligned with the temperature. This deviation arises from the gray-body factors, discussed in Sec.~\ref{sec:Hawking}.

In our analysis, we used the PBH temperature $T_{\text{PBH}}$ to estimate the time window $\tau_B$ during which a given experiment would be sensitive to the PBH burst. As shown in Fig.~\ref{fig: 10s1day}, this is a good approximation, since integrating the signal from earlier times has little impact on the final result. However, this does not imply that particles with energy $E \sim T_{\text{PBH}}$ are emitted exclusively within the last $\tau_B$ seconds of the burst. For example, Fig.~\ref{fig: InstSpec} shows that particles with energies around $4 \times 10^3~\GeV$ begin to be emitted roughly a day before the PBH fully evaporates. However, their flux remains negligible until the PBH temperature reaches $\sim 4 \times 10^3~\GeV$, which occurs about 10 seconds before it fully evaporates. Despite the three orders of magnitude difference in time between, the total number of particles emitted above this energy threshold changes very little. This is because, at such high energies, the particle flux is overwhelmingly dominated by the final stages of evaporation.  

\begin{figure}[t]
    \centering
    \includegraphics[width=1\linewidth]{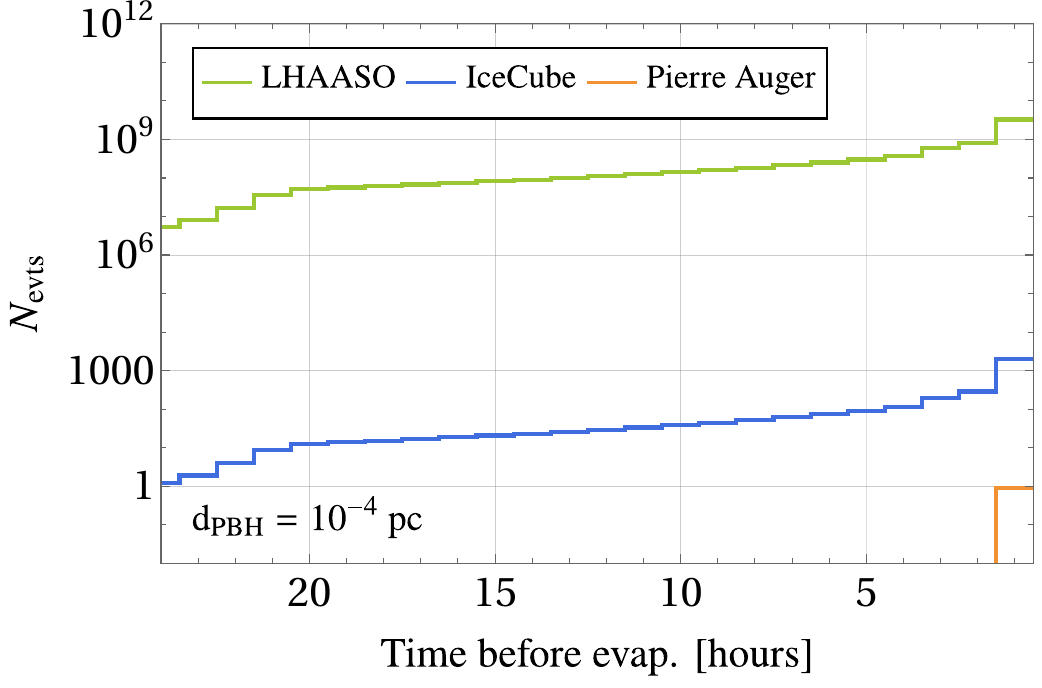}
    \caption{Expected number of events at Earth in one day before the full evaporation of a PBH located at declination $\delta = -0.02^{\circ}$, assuming a distance of $4 \times 10^{-4}$ pc from Earth. We consider photons arriving at LHAASO (green) and Pierre Auger (orange), as well as neutrinos arriving at IceCube (blue).}
    \label{fig: TotalEvtsDiffExps}
\end{figure}

Having discussed how associating the PBH temperature with the peak energy of the emitted particles helps identify the time window of maximum experimental reach, we now turn to the background rate and examine how it limits the sensitivity, thereby limiting the maximum PBH distance each experiment can probe. Fig.~\ref{fig: NevtsRate} shows the differential event rate $dN_{\rm evt}^{\nu_\mu}/dt$ (see Eqs.~\eqref{eq:dNevtdt_nu_primary} and \eqref{eq:dNevtdt_nu_secondary}) for IceCube (left panel) and LHAASO (right panel), for varying PBH distances and including the relevant background rate. As a case example, we compute the expected event rate, and total number of events binned in hours, for IceCube, LHAASO and Pierre Auger for a PBH in a specific declination, chosen as $\delta = -0.02^{\circ}$. For IceCube, such declination means that the neutrinos would arrive from above the detector. At this coordinate, the muon background from cosmic ray interactions in the atmosphere ~\cite{IceCube:2015wro} significantly limits the time window during which IceCube could detect a PBH evaporation signal. 
Muon tracks can mimic the long-track signatures of muon neutrino interactions, making them a dominant background.
Specifically, for $d_{\text{PBH}} = 10^{-4}$~pc only the final $\mathcal{O}(100~\text{s})$ are effectively background free.
It is important to emphasize that the muon background strongly depends on the declination. For trajectories coming from below the detector, atmospheric muons are strongly attenuated by the large column density of the Earth, making it extremely unlikely for them to reach IceCube. In contrast, atmospheric muon neutrinos can traverse the Earth and still interact in the detector. Both types of background for the chosen declination are shown in Fig.~\ref{fig: NevtsRate}. 

For photons, we use LHAASO, which energy range covers $\sim 100$ GeV -- PeV, as a representative example. Using the effective area of the water Cherenkov detector array~\cite{Yang:2024vij} (WCDA), particularly at the same declination considered before, $\delta = -0.02^{\circ}$, and computing the cosmic ray background following Refs.~\cite{Yang:2024vij,Abdo:2014apa}, we find that LHAASO could be sensitive to PBH signals up to several days before the final burst, reflecting the comparatively more favorable detection conditions for photons, primarily due to their much larger electromagnetic interaction cross-section compared to the weak interaction cross-section of neutrinos.

For higher energy threshold telescope, as Pierre Auger with $E_{\rm threshold}\sim 3\times 10^9$ GeV, only the very last instant of the evaporation is going to be relevant, since higher energy particles are emitted only during this final moment. As a consequence, even though the event rate can be high during this moment, because of the tiny fraction in which this spike occur, the total number of events is completely negligible compared to other telescopes in the range of GeV to PeV. In particular, Pierre Auger can achieve an event rate of order $\sim 10^{25}$ sec$^{-1}$ but just over a period of less than $10^{-25}$ sec, resulting in less than one event over the lifetime. We illustrate this behavior by integrating the event rate over one hour beans for the interval of a day, as shown in Fig.~\ref{fig: TotalEvtsDiffExps}.

\vspace{1em}

\section{Triangulation}\label{app:triang}

%%%%%
\begin{figure*}[t]
    \centering
    \includegraphics[width=0.95\textwidth]{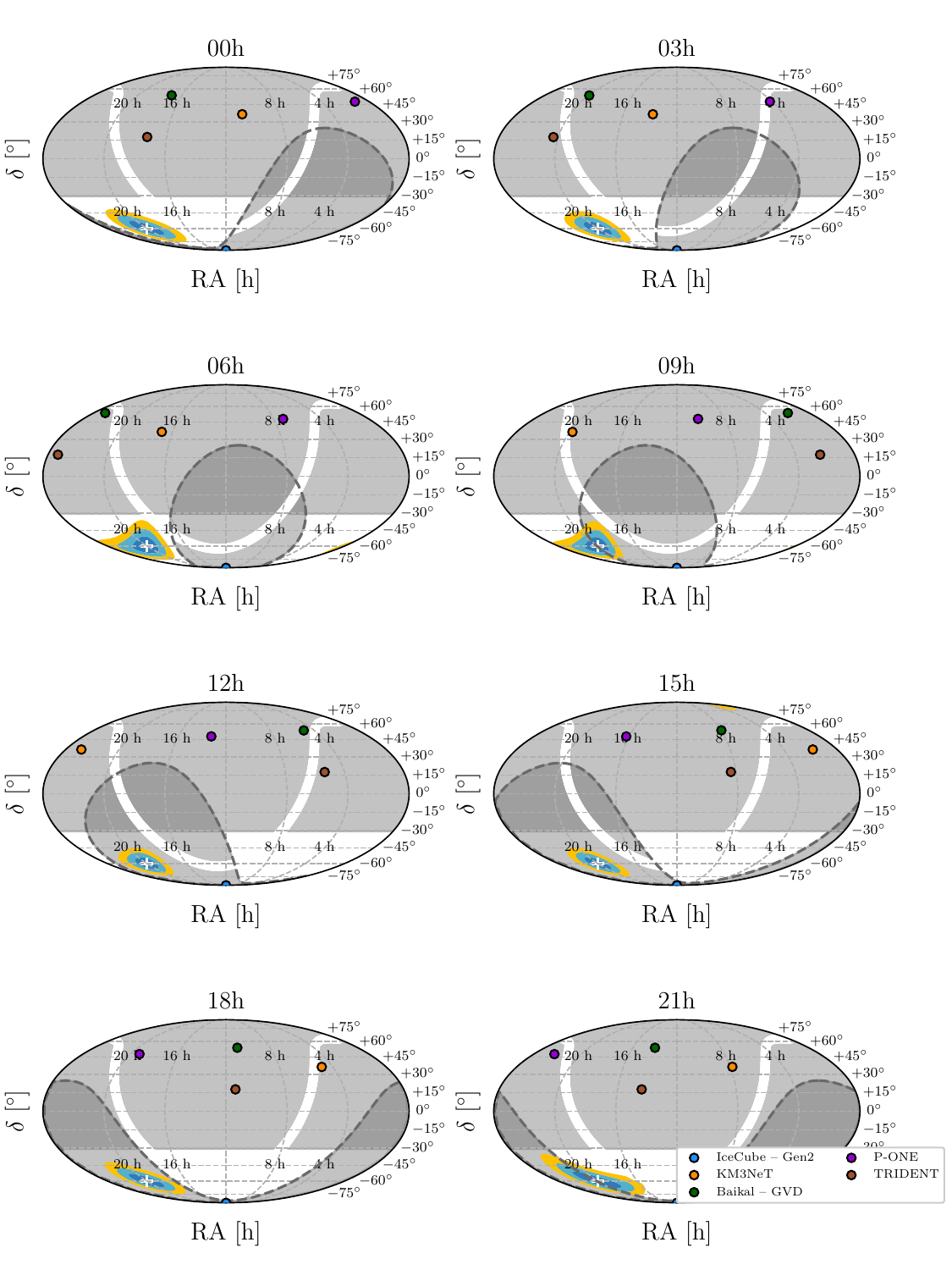}
    \caption{The reconstruction of the PBH position from the combined measurement of the PBH burst across several neutrino detectors. We show the 1~$\sigma$ region around the true PBH position $\delta_{\text{PBH}} = -60^\circ$, $\text{RA}_\text{PBH}= 300^\circ$ (20~h), for several detection times, varying during the day of the benchmark event. The detectors used in this analysis are listed in Tab.~\ref{tab:Experiments_Location} except IceCube. The upper gray region is the one-day coverage by LHAASO and HAWC, while the region with the dashed contour is the instantaneous field of view of Pierre Auger at each hour on 13th of February 01:00:00 UTC. }
    \label{fig:triangulation_pbh}
\end{figure*}
%%%
Neutrino detectors may also provide sufficient directional information to localize the source in the sky. In Fig.~\ref{fig:triangulation_pbh} we show the results for the PBH triangulation for an event located at $\delta_{\text{PBH}} = -60^\circ$ and $\text{RA}_\text{PBH}=300^\circ$ (20h). The position reconstruction is based on the difference in arrival times of the neutrinos in the different neutrino telescopes, following the same methods as in Ref.~\cite{Muhlbeier:2013gwa}. Colored contours denote $1\sigma$ CL (yellow), $2\sigma$ CL (light blue) and $3\sigma$ CL (dark blue) in the reconstruction of the PBH location. We show results for the benchmark date event and vary the burst detection hour. The dependence on the hour of the day comes from the relative position of the event and the different experiments; that is, it varies as the instantaneous effective area. This speaks to the need for having multiple neutrino telescopes covering different portions of the sky simultaneously. The minimum requirement is to have at least four operating detectors, such that we break all degeneracies in determining the PBH position. 
For this example, we are considering the neutrino telescopes given in Tab.~\ref{tab:Experiments_Location} except IceCube.

We note however that the direct measurement of tracks in current and future neutrino telescopes surpass the determination of the sky's position of the neutrino source. Thus, the triangulation considered here can be considered as a cross-check to direct observations.

\bibliography{transient_sources}% Produces the bibliography via BibTeX.

\end{document}